\documentclass[useAMS,usenatbib]{mn2e}

\usepackage{graphicx}
\usepackage{epsf}

\newcommand{\Om}{\Omega}

\newcommand{\Si}{\Sigma}
\newcommand{\si}{\sigma}

\newcommand{\Cov}{{{\rm Cov}}}
\newcommand{\Mpc}{{\ensuremath{\rm Mpc}}}
\newcommand{\Gpc}{{\ensuremath{\rm Gpc}}}

\newcommand{\FoM}{{\ensuremath{\rm FoM}}}

\title[BAO for LAMOST]{Forecasting the Dark Energy Measurement with
 Baryon Acoustic Oscillations: Prospects for the LAMOST surveys}

\author[X. Wang et al.]{
\parbox{\textwidth}{
Xin Wang$^1$, Xuelei Chen$^{1*}$, Zheng Zheng$^{2,3}$, Fengquan Wu$^1$, 
 Pengjie Zhang$^4$ and Yongheng Zhao$^1$
}\vspace*{4pt}\\
$^1$ National Astronomical Observatories, Chinese Academy of Sciences,
20 A Datun Road, Chaoyang District, Beijing, 100012, China.\\
$^2$ John Bahcall Fellow\\
$^3$ Institute for Advanced Study, Einstein Drive, Princeton, NJ08540, USA\\
$^4$ Shanghai Astronomical Observatory, Chinese Academy of Science,
80 Nandan Road, Shanghai 200030, China\\
$^*$ xuelei$~@~$cosmology.bao.ac.cn
}

\begin{document}

\date{Accepted . Received ;}

\pagerange{\pageref{firstpage}--\pageref{lastpage}} \pubyear{2007}

\maketitle

\label{firstpage}

\begin{abstract}
The Large Area Multi-Object Fiber Spectroscopic Telescope (LAMOST) is a 
dedicated spectroscopic survey telescope being built in China, with an 
effective aperture of 4 meters and equiped with 4000 fibers.
Using the LAMOST telescope, one could make redshift survey of
the large scale structure (LSS). The baryon acoustic oscillation (BAO)  
features in the LSS power spectrum provide standard rulers 
for measuring dark energy and other cosmological
parameters. In this paper we investigate the meaurement precision 
achievable for a few possible surveys: 
(1) a magnitude limited survey of all galaxies, 
(2) a survey of color selected red luminous galaxies (LRG), 
and (3) a magnitude limited, high density survey of $z<2$ quasars. 
For each survey, we use the halo model to estimate the bias of the 
sample, and calculate the effective volume. We then use the
Fisher matrix method to forecast the error on the dark energy 
equation of state and other cosmological parameters for different 
survey parameters. In a few cases we also use the Markov Chain Monte Carlo
(MCMC) method to make the same forecast as a comparison. The fiber time 
required for each of these surveys is also estimated. 
These results would be useful in designing the surveys for LAMOST. 
\end{abstract}


\section{Introduction}

Baryon acoustic oscillations (BAO) before recombination left wiggling 
features in the matter power spectrum. At large scale where the
evolution of the density perturbation is still linear, such features
is preserved in the galaxy power spectrum \citep{EH98,M99,EH97}. 
This provides us with a well calibrated standard ruler, 
which enables precise measurement of cosmological parameters, 
especially the dark energy parameters
\citep{EW04,E05a,SE03,SE07,K06,L03,GB07,BG03,B06}.
Like cosmic microwave background (CMB), the physics involved is relatively 
clean and well understood, hence it is arguably less 
affected by unknown systematic errors, which might undermine empirical-rule  
based methods such as the type Ia supernovae (SNIa) and the cluster abundance 
measurements. 

Recently, such a BAO feature have been observed in the 2dF galaxy 
redshift survey (2dFGRS) \citep{2dF05} and the
Sloan Digital Sky Survey (SDSS) survey data\citep{E05b,H05a,H05b,T06,P07b,O07,
CG08a,CG08b}. 
A number of more powerful BAO surveys are being planned.  
Some examples are WiggleZ \citep{G07}, SDSS-3 (BOSS)
\footnote{{\it http://cosmology.lbl.gov/BOSS/, http://www.sdss3.org/ }}, 
HETDEX \footnote{{\it http://www.as.utexas.edu/hetdex/}}, 
WFMOS \citep{BNE05}, ADEPT \footnote{{\it http://www7.nationalacademies.org/ssb/be\_nov\_2006\_bennett.pdf}}, Space/Euclid \citep{C08} 
in the optical/IR and FAST 
\footnote{{\it http://www.bao.ac.cn/LT/}}, HSHS\citep{PBP06}, 
SKA \citep{AR05} \footnote{{\it http://www.skatelescope.org/}} in the radio. 

In this paper, we make forecasts on cosmological constraints from
BAO measurement in future surveys with the
{\it Large Sky Area Multi-Object Spectroscopic Telescope}, or LAMOST
\footnote{{\it http://www.lamost.org/}}\citep{C98,R08}, 
a telescope being built in
China. The LAMOST is a 4 meter 
Schmidt telescope with a field of view of 20 $\deg^2$.
Equipped with 4000 optical fibers that are individually positioned
by computer during observation, it is able to take spectra of 4000 targets 
simultanously. More details of the LAMOST telescope are described
 in Appendix A. 
The LAMOST is ideally suited for conducting large 
scale redshift surveys.

We expect that once LAMOST becomes operational, it will carry out 
several different survey projects. The design of such surveys should 
maximize the scientific output for a given amount of observing 
fiber time. Some obvious choices include a magnitude limited 
general galaxy redshift survey, and a magnitude limited quasar survey.  
For the purpose of BAO measurements, it is desirable to 
have additional surveys for targets at relatively higher redshift.
Often found to be the bright central galaxies of galaxy groups and clusters,
the luminous red galaxies (LRG) \citep{E01,BD03,Z05} can be detected 
at higher redshifts than a typical galaxy for a given apparent magnitude.
Quasars can be observed at even higher redshifts. 
In the SDSS surveys, quasars were very sparse, making it
difficult to use in BAO measurement \citep{S05,S07}.
However, it is expected that the density of observable quasars with the 
LAMOST is much higher, since a lower limiting flux can be reached.
Therefore, we also consider quasar samples in our study, with quasar 
themselves as the tracer of the large-scale structure (LSS).
The absorption lines (Ly$\alpha$ forest) in the quasar spectra also
provide useful information \citep{W03,M06,ME07}, 
which will be considered in our future work. 
We describe these surveys in more
detail (including the selection of the sample, the required observing 
fiber hours, and the estimate of bias of the sample) in Appendix B.

We will make forecasts of the measurement errors on the dark energy equation
of state and other cosmological 
parameters with these surveys, and also estimate the resources required for 
these surveys. We will first use the Fisher matrix method \citep{SE03,SE07}. 
Then for a few cases we also use a Markov Chain Monte Carlo (MCMC) simulation
to make the forecast. The advantages of the Fisher matrix method are that 
it can easily be used to explore large parameter space and that the 
relationship between the input and output is very clear. 
The MCMC method does not assume
the likelihood to be Gaussian and thus can probe the full shape of the 
likelihood surface.

Previous studies have shown that it is feasible with the LAMOST to obtain 
better precisions in cosmological parameters than ongoing surveys like 
the SDSS \citep{FCY00,SSF06,LXFZ08}.
The study by \cite{FCY00} were conducted before the importance of BAO 
in dark energy measurement was widely recognized. \cite{SSF06} considered 
the Alcock-Paczynski test \citep{AP79} in real space \citep{MS03}, with a 
method different from the one we use.  In all of these studies, 
a single survey of galaxies up to $B=20.5$ were considered, assuming  
an total survey area of $15000 \deg^2$, a total number of galaxies
$10^7$, and a galaxy sample bias of 1. 

In this paper, we try to make a {\it realistic} assessment of LAMOST
surveys. We consider different samples that could be collected by the LAMOST
telescope within a reasonable amount of observation time. The surveyed 
sky area is 
assumed to be about $8000 \deg^2$, for which the SDSS photometric survey 
catalogue is currently available for target selection. We use the published 
SDSS luminosity function to estimate the number of targets.
In order to accurately estimate the measurement error for the given sample, 
it is necessary to know the bias of the sample. We use the halo model to
estimate the bias of the samples. 

It should be noted that in the methods employed here
\citep{SE03, SE07}, the treatment of the non-linear effect 
and redshift distortion is very crude. Furthermore, the assumption of scale
independent clustering bias may be overly 
idealistic. Recently, 
\cite{CS07} studied the BAO in both the power spectrum and 
correlation function by using the renormalized perturbation
theory(RPT, \cite{CS06a,CS06b}). They showed that mode-coupling 
typically leads to percent-level shifts in the acoustic peak of 
correlation function. \cite{SSS08} have also shown with 
both numerical simulation and analytical calculation 
that the peak is moved by these effects.
\cite{A08} proposed a new method to extract the unbiased estimation
of sound horizon scale based on a large N-body simulation 
and semi-analytical galaxy formation model.
\cite{S08} also presented high signal-to-noise ratio measurements 
of acoustic scale from large volume simulations, and obtained a robust 
measurements of acoustic scale with scatter close to that 
predicted by the methods utilized in this paper.
There have also been works on scale dependent halo and galaxy bias
\citep{SSS07}. 
In analysing the real data to extract cosmological information,
a more refined treatment is required to account for all 
these effects.

We review the methods for error forecasting in \S~2. \S~2.1 is devoted to the 
Fisher matrix method, which was developed by \cite{SE03,SE07} 
for BAO measurements; \S~2.2 is devoted to the MCMC method;
in \S~2.3 we discuss how to estimate the bias of the sample for the planned 
surveys. After a brief description of the data that 
we assume LAMOST could collect, 
we present our forecast on the BAO measurement precision in \S~3, 
and conclude in \S~4. In the appendices, we describe the 
characteristics of the LAMOST telescope (Appendix A),
and the details of our survey design (Appendix B), 
including the generic main survey (Appendix B1), 
the LRG survey (Appendix B2), and the QSO survey (Appendix B3). 
We estimate the density and bias of each sample, 
and also discuss the fiber time required for completing each survey.

Throughout the paper, we adopt the flat $\Lambda$CDM, with the WMAP three 
year \citep{SB07} best 
fitted parameter values (~$\Om_bh^2=0.0223, ~ \Om_mh^2=0.128, ~
\Om_k=0,~ h=0.73,~ w_0=-1,~ w_a=0,~ n_s=0.958,~ A_S=2.3 \times
10^{-9},~ \tau=0.092$~) as our fiducial model.


\section{Methods of Forecast}
\subsection{Fisher Information Matrix}
Before recombination, the presence of large number of free electrons
ensures that the photons and baryon plasma are tightly coupled. This
provides a resilient force against any 
motion induced by gravity. Acoustic waves are generated 
from the primordial fluctuations. When the Universe is sufficiently
cooled for the photons and baryons to decouple, the oscillation
ceases and the waves are imprinted on the matter and radiation 
density power spectra \citep{HW96,EH97,EH98,M99}, with the (comoving)
characteristic scale (the BAO scale) determined by the sound horizon at the 
last scattering surface:
\begin{eqnarray}
  s &=& \int_0^{t_{\rm rec}}\,c_s\, (1+z)dt
     = \int_{z_{\rm rec}}^\infty {c_s\;dz\over H(z)} ,
\label{eq:rsound}
\end{eqnarray}
where $c_s$ is the sound speed. 
For a given set of cosmological parameters, 
the absolute scale of sound horizon can be calculated. 

When making measurement of the matter power spectrum in a galaxy
survey, the distances along and perpendicular to the line of sight are
measured from redshift and angular separation respectively. The
comoving distances are given by \citep{SE03}
\begin{eqnarray}
r_{\perp} &=&(1+z)D_A(z)\Delta \theta,  \nonumber \\
r_{\parallel} &=&\frac{ c \Delta z}{ H(z) }.
\label{eqn:staRuLer}
\end{eqnarray}
The BAO scale provides a standard ruler to
measure the angular diameter distance $D_A(z)$ and the Hubble
expansion rate $H(z)$. In a model with dark energy equation of state
$w(z)$, these are given by 
\begin{equation}
\label{eqa:HubZ}
\frac{H(z)}{H_0}=\left[\Om_m(1+z)^3+  \Om_k(1+z)^2 +
\Om_X e^{3\int_0^z\frac{1+w(z)}{1+z}dz} \right]^{1/2}
\end{equation}
and 
\begin{equation}
\label{eqa:DaZ}
D_A(z)=\frac{c}{1+z}\int_0^z\frac{dz}{H(z)} .
\end{equation}
In this paper, we consider both a constant equation of state $w$ for the 
dark energy, and a redshift-dependent one parameterized
in the form of
\begin{equation}
\label{eqa:eos}
w(z)=w_0+w_a [1-a(z) ]=w_0+w_a\frac{z}{1+z}.
\end{equation}

The statistical error in LSS measurements 
includes sampling variance due to the finite volume of the survey, as
well as shot noise. In Fourier space, the statistical error 
is the summation of these two effects \citep{FKP94,T97}:
\begin{eqnarray}
\label{eqa:errBP}
\frac{\si_P}{P}=2\pi \sqrt{\frac{1}{V_s k^2 \Delta k }}
 \frac{P(k)+1/n}{P(k)}= 2\pi \sqrt{\frac{1}{V_{\rm eff}(k) k^2 \Delta k }},
\end{eqnarray} 
where $V_{\rm s}$ is the survey volume and $V_{\rm eff}$ is the effective 
volume defined as 
\begin{equation}
\label{eqa:Veff}
V_{\rm eff}(k,\mu) =  \int \left[\frac{n(\vec r)P(k,\mu)}
{n(\vec r)P(k,\mu)+1}\right]^2 d{\vec r}.
\end{equation}
Here $\mu$ is the cosine of the angle between direction of 
$\vec{k}$ and the line of sight, and $n(\vec r)$ is the comoving
number density of galaxies. For constant $n$,
\begin{equation}
\label{eqa:Veff1}
V_{\rm eff}(k)= \left[\frac{nP(k)}{nP(k)+1}\right]^2 V_{\rm s}.
\end{equation}

The Fisher information matrix for cosmological parameters derived from
an LSS measurement is given by \citep{T97, SE03}
\begin{equation}
\label{eqa:Fishij}
F_{ij}=\int_{\vec{k}_{\rm min}}^{\vec k_{\rm max}}\frac{\partial 
\ln P(\vec k)}{\partial \it{p}_i}\frac{\partial \ln P(\vec k)}
{\partial \it{p}_j}V_{\rm eff}({\vec k}) \frac{d\vec k}{2(2\pi)^3},
\end{equation}
where $p_i$ denotes the cosmological parameters in the theory,

In reality, the galaxy power spectrum is observed in
redshift space. The conversion from redshifts to physical scales depends
on cosmological parameters \citep{SE03}, 
\begin{eqnarray}
\label{eqa:Pobs}
P_{\rm obs}(k_{\rm ref\perp},k_{\rm ref\parallel})&=&\frac
{D_A(z)_{\rm ref}^{2}H(z)}{D_A(z)^{2}
H_{\rm ref}(z)}\left(1+\beta\frac{k_{\parallel}^{2}}{k_{\perp}^{2}
+k_{\parallel}^{2}}\right)^{2} \nonumber\\
& &\times b^{2}G(z)^{2}P_{\rm m0}(k)+P_{\rm shot},
\end{eqnarray}
where the subscript ``ref'' denotes quantities calculated in the 
reference cosmology, 
$b$ is the bias factor for the galaxy sample, $G(z)$ is the growth
factor, and $\beta$ is the 
redshift distortion factor, $k_{\perp}$ and $k_{\parallel}$ are the $k$
components perpendicular and parallel to the line of sight, respectively, and
$P_{\rm shot}$ is the shot noise contribution. 

The measurement error of the power spectrum depends on the amplitude of
the power spectrum, which is given by $P_{\rm obs}(k)=b^2 P_{\rm lin}(k)$ 
in the 
$k$ range relevant for BAO measurement. Thus, 
the uncertainty depends sensitively on the value of the 
bias $b$ of the sample. We present our method of estimating the bias
parameter in \S~2.3.

The measurements are also subject to some
systematic errors, such as the effect of redshift-distortion due to 
peculiar velocity, non-linear evolution of the power spectrum,  
scale-dependent bias, and finite spectrograph resolution.

Non-linear evolution of the density fluctuation 
enhances small scale power,
which corresponds to a smearing of the acoustic 
signature \citep{ESW06,SE07}. This can be 
estimated by convolving a Gaussian displacement field with the 
linear correlation function,
\begin{eqnarray}
\label{eqa:Pnonl}
P_{\rm nonl}(k,\mu)=P_{\rm lin}(k,\mu)\exp\left(-\frac{{k_{\perp}^{2}}
{\Si_{\perp}^{2}}}{2}- \frac{{k_{\parallel}^{2}}{\Si_{
\parallel}^{2}}}{2}\right)  
\end{eqnarray}
where 
$$\Si_{\perp}=\Si_{\it 0}G, \qquad \Si_{\parallel}=\Si_{\it 0}G(1+f),$$ 
with $\Si_0=12.4h^{-1}\Mpc$ for $\sigma_8=0.9$, and $f=d\ln G/d\ln a$ is 
the derivative of the growth factor. 
The observed power spectrum is smeared below the resolution of
the spectroscopic measurement. We model such smearing as a Gaussian, 
\begin{eqnarray}
\label{eqa:photo}
P(k,\mu)=P_{\rm obs}(k,\mu)\exp\left(-k_{\parallel}^2 \si_{\it r}^2
           \right)  
\end{eqnarray}
    where $\si_{\it r} = c\si_{\it z}/H(z)$, and $\si_{\it z}=(1+z)
 \si_{0}$, $\si_{0}=1/R $, with $R$ the resolution of the spectroscopic 
measurement. We consider $R=1000$ and $R=2000$ respectively (for more 
details see Appendix A). The finite spectra resolution has the effect
of smearing small scale powers. We set $k_{\rm max}=0.5 h/\Mpc$.
The modes at very large scales contribute very little to the 
integral in equation~(9), so we set 
$k_{min}=0$. 

The parameters involved include cosmological
parameters and survey parameters. To define a cosmological model, 
the following set of parameters are included: the Hubble 
constant $h$, baryon density $\Omega_b$, 
matter density $\Omega_m$, dark energy density $\Omega_X$, dark energy 
equation of state $w$,  spectrum normalization $\ln A_s$, 
spectral index $n_s$, reionization optical depth $\tau$. In addition, 
for each redshift bin $z_i$ of the survey, we have parameters $\ln H(z_i)$,
$\ln D_A(z_i)$, growth function $G(z_i)$, linear redshift distortion
$\ln \beta(z_i)$, and shot noise $P_{shot}(z_i)$. 

To obtain useful constraints on cosmological parameters, it is necessary to 
break the degeneracy by combining the BAO data with data obtained from 
some other cosmological observations, e.g., CMB and/or Type Ia supernovae 
(SNIa). 
Consider first the combination of BAO and CMB, 
the total Fisher matrix is then given by
\begin{equation}
\label{eqa:Fadd}
F^{tot}_{ij}= F^{CMB}_{ij}+\sum_{n} F^{LSS}_{ij}(z_n),
\end{equation}
where $ F^{LSS}_{ij}(z_n)$ is the Fisher matrix derived from the $n$th
redshift bin of the large scale structure redshift survey. The CMB Fisher 
matrix is expressed as 
\begin{equation}
\label{eqa:fisherCMB}
F^{CMB}_{ij}=\sum_{l}\sum_{X,Y}\frac{\partial C_{Xl} }
        { \partial p_{\it i}} (\Cov)_{l,XY}^{-1} \frac
        {\partial C_{Yl} }{\partial p_{\it j} },
\end{equation}
where $C_{Xl}$ is the $l^{th}$ multipole for $X=T$ (temperature correlation), 
$E$ ($E$ mode polarization correlation), $B$ ($B$ mode polarization 
correlation), and $C$ (temperature-polarization cross-correlation), 
respectively \citep{SZ96,KKS96,S97,EHT99}. 
The elements of the covariance matrix $\Cov_l$ between various
power spectra are
\begin{eqnarray}
(\Cov)_{l,TT} =\frac{2}{(2l+1)f_{\rm sky}} (C_{Tl} + \omega_T^{-1}
 B_l^{-2}  )^2 ,\\
(\Cov)_{l,EE} =\frac{2}{(2l+1)f_{\rm sky}} (C_{El} + \omega_P^{-1}
 B_l^{-2}  )^2 ,\\
(\Cov)_{l,BB} =\frac{2}{(2l+1)f_{\rm sky}} (C_{Bl} + \omega_P^{-1}
 B_l^{-2}  )^2 ,\\
(\Cov)_{l,CC} =\frac{2}{(2l+1)f_{\rm sky}} (C_{Cl}^2 + 
(C_{Tl} + \omega_T^{-1} B_l^{-2}  ) \nonumber\\ \times (C_{El} + 
\omega_P^{-1} B_l^{-2}  ),\\
(\Cov)_{l,TE} =\frac{2}{(2l+1)f_{\rm sky}} C_{Cl}^2  ,\\
(\Cov)_{l,TC} =\frac{2}{(2l+1)f_{\rm sky}} C_{Cl}(C_{Tl} +\omega_T^{-1}
 B_l^{-2}  ) ,\\
(\Cov)_{l,EC} =\frac{2}{(2l+1)f_{\rm sky}} C_{Cl}(C_{El} +\omega_P^{-1}
 B_l^{-2}  ) ,\\
(\Cov)_{l,TB} = (\Cov_l)_{EB} = (\Cov_l)_{CB} =0,
\end{eqnarray}
where  $\omega_T, \omega_P$ are the inverse square of the detector
noise level on a steradian patch for temperature and polarization, 
respectively. $B_l^2=\exp[-l(l+1)\theta^2_{\rm beam}/8\ln2]$ is the 
beam function, $\theta_{\rm beam}$ is the full-width, 
half-maximum (FWHM) of the beam in radians.
We shall assume that the CMB data would come from Planck\footnote{{\it http://www.rssd.esa.int/index.php?project=planck}}, which is scheduled to be launched 
in 2009. By the time that 
the data of any of the surveys considered here is collected, Planck should 
have been in operation for at least three years. Therefore, we assume that
CMB measurement errors correspond to Planck three-year observation (see 
Table~ \ref{tab:Planck}).

\begin{table}
\caption{\label{tab:Planck}Assumed Planck Characteristic with three year 
integration.}
\begin{tabular}{l c c c }
\hline
Center Frequency (GHz)  &     100  & 143 & 217 \\ 
\hline
Angular Resolution (arcmin) & 10.0 & 7.1 & 5.0   \\
$\Delta_T$ per pixel $(\mu K ) $    &   3.9 & 3.5 & 7.6 \\ 
$\Delta_P$ per pixel $(\mu K) $    &   6.3  & 6.6 & 15.4  \\
sky fraction $f_{\rm sky}$ &  0.65 & 0.65 &0.65 \\
\hline
\end{tabular} 
\end{table}

In summary, similar to \cite{SE03}, we consider the parameter set,
\begin{equation}
\{p_{i,CMB } \}=\{\Omega_mh^2, \Omega_bh^2, \Omega_m,  n_s, \ln A_s, \tau,
 D_{A,CMB}\}, 
\end{equation}
and the following additional parameters for LSS:
\begin{eqnarray*}
\{p_{i,LSS}  \}&=&\{
\ln D_{A,i},\ln H_i,\ln G_i, \ln\beta_i, P_{shot,i}\}.  
\end{eqnarray*}
We then marginalize the nuisance parameters 
$( \ln G_i, \ln\beta_i, P_{shot,i})$
by selecting the submatrix of $F^{-1}_{ij}$ with appropriate column and rows. 
Finally, the Fisher matrix of dark energy
parameters $w_0$ and $w_a$ was obtained by converting from the distance 
parameter space
$\{p_m\} =\{Da_i,~ H_i\}$ to the dark Energy parameter space 
$\{q_i\} =\{ w_0,~w_a,~\Omega_X \}$.
\begin{equation}
F^{DE}_{ij}=\sum_{m,n}\frac{\partial {\it p}_m}{\partial {\it
    q}_i}F^{dis}_{mn} \frac{\partial
 {\it p}_n}{\partial {\it q}_j}.
\end{equation}

\subsection{Monte Carlo Simulation}

Another method of making forecast is to use Monte Carlo to analyse 
a synthetic data set (e.g., power spectrum) with the expected error for
the experiment in consideration \citep{PL06, LXFZ08}. 
This method is more time consuming than the Fisher matrix method, 
but it does not assume the likelihood to be nearly Gaussian.

In order to deal with dynamical dark energy model, we modified the Boltzmann 
code ${\tt camb} $ together with Monte Carlo program 
${\tt CosmoMC} $\footnote{{\it http://cosmologist.info/cosmomc/}} \citep{LB02}
to search through the cosmological parameter space \{$\Omega_bh^2, 
\Omega_ch^2, \theta, \tau, w_0, w_a, n_s,  A_s , ( f_{\nu}, n_r)$  \},
where $\theta$ is the
angle extended by the sound horizon at recombination and serves as an 
independent variable in lieu of the Hubble constant here. 
We generate the matter power 
spectrum and the CMB temperature and polarization angular power spectra.
Again we assume Planck three-year (Table. \ref{tab:Planck}) 
data that is generated by the program ${\tt FutureCMB}$ 
\footnote{{\it http://lappweb.in2p3.fr/$\sim$perotto/FUTURCMB/home.html}}
\citep{PL06}. We do not include lensing effect here, since it has no 
significant impact on our results but requires more computer time.

For the LSS part, we assume that the uncorrelated band power is a 
Gaussian realization with band 
error $(\sigma_P/ P)^2 \sim 1/ (k^2 V_{\rm eff}\Delta k ) $. 
The bin size of $k$ is chosen to be similar to the SDSS-II LRG real data 
at low redshifts. The mock data include larger range of $k$ for 
higher redshifts to account for the fact that the linear approximation is 
valid at smaller scales then. The bias of the sample is also treated
as a parameter in the MCMC.

The overall constraint on dark energy could be improved significantly by 
adding data from complementary observations.
As we are interested in 
estimating what could be learned from ``clean'' signals such as BAO, and
also what could be achieved with all types of experiments,
we consider the constraints from two cases:
(1) only the CMB and BAO data; (2) CMB, BAO and SNIa data.
In the second case, we assume that a large number of SNIa (1200 in total at 
$0<z<1$) collected by SNLS \citep{SNLS} are available.

\subsection{Estimation of the Bias Factor}
To make an estimate of the measurement with either
the Fisher matrix or the MCMC method, one needs to know the 
bias of the sample, which determines the amplitude of the power spectrum.
With the real data, this can be measured for the assumed cosmological model
\citep{C05,R06,BCL07,R07,P08},
 but in planning the survey
one needs some way to estimate the value of the bias factor. Here we use the 
halo model to make such an estimate.

In the halo model \citep{J98,S00,P00,SS01,SD01,CS02},
the number density of galaxies more luminous than a threshold
luminosity $L_f$ is
\begin{equation}
\bar{n}_g= \int_{0}^{\infty}dm\frac{dn}{dm}
\langle  N(m,>L_f) \rangle,
\end{equation}
and the large scale galaxy bias factor is 
\begin{equation}
\label{eqa:bg}
b_g=\frac{1}{\bar{n}_g}\int_{0}^{\infty}dm
\frac{dn}{dm} \langle N(m,>L_f)\rangle b(m). 
\end{equation}
In the above expressions, 
$dn/dm$ is the halo mass distribution function, and $b(m)$ is the halo
bias factor, for which we use the fitting formulae in 
\cite{ST99}. 
The function $\langle N(m,>L)\rangle$ represents the 
halo occupation distribution function (HOD) for 
galaxies above luminosity $L$ in halos of mass $m$, which can be 
parameterized as a sum of contributions from central and
satellite galaxies \citep{ZZW05,ZZ05,BW02}.
In our calculation of the galaxy number densities and 
large-scale bias factors, we adopt HOD parameters determined 
from existing galaxy clustering data (see Appendix B).

For the large-scale bias factor of QSOs, we follow the model
in \citet{WL03} (also see \citealt{MC06}). In this
model, the QSO activity is triggered by halo-halo major 
mergers. The black hole powering the QSO is assumed to be
a fraction 
of the mass of the host halo. The QSO shines at
the Eddington luminosity after the merger with a lifetime 
equal to the dynamical time of a galactic disk, with a universal 
light curve. 
The model
gives a deterministic relation between the $B$-band 
luminosity $L_B$ and the halo mass $m$ at a given redshift 
$z$ and predicts the $B$-band luminosity function 
$\Psi(L_B,z)$. 
The bias factor for quasars more luminous than $L_{B,min}$
is given by
\begin{equation}
b(z) = \frac{\int_{L_{B,min}}^{\infty} b(L_B(m,z )) 
       \Psi(L_B, z) dL_B}
        {\int_{L_{B,min}}^{\infty} \Psi(L_B, z) dL_B  }.
\end{equation}
The model is not necessarily realistic, but for our purpose
in this paper, we find that the uncertainty in the estimated bias from this 
simple model does not affect our results too much (see \S~3.2).


\section{Results}

\begin{table}
\caption{\label{tab:summary}Summary of surveys. }
\begin{tabular}{ccccc}
\hline
Sample &Magnitude & Surface Density  & Targets  & Fiber Hours \\
       &Limit     & ($\deg^{-2}$)    & ($10^6$) & ($10^6$)\\
\hline
MAIN1 & $r<18.8$ & 330 & 2.6  & 0.88   \\
MAIN2 & $r<19.8$ & 1050 & 8.4  & 14    \\
LRG & $i_{dev}<20$ & 205 & 1.5  & 1.9  \\
QSO1 & $g<20.5$  & 30 & 0.24  & 0.06   \\
QSO2 &$g<21$  & 45 & 0.36  & 0.18      \\
QSO3 &$g<21.65$ & 72 & 0.57  & 0.855     \\
\hline
\end{tabular}
\end{table}

We shall consider three surveys for the 
LAMOST telescope: (1) a magnitude-limited general 
survey of galaxies of all types which we shall call it the {\it main survey};
(2) an LRG survey; (3) a magnitude-limited low redshift quasar 
survey ($z<2.1$).
The characteristics of the LAMOST telescope 
and the details of the sample construction and 
survey design are discussed in the Appendices.

Table ~\ref{tab:summary} gives
a summary of these surveys, including the magnitude limit, the surface density, 
the total number of targets, and the required fiber hours, i.e., the number of 
fibers allocated for the survey multiplied by the observing hours.
For the main and QSO surveys, several different magnitude limits are 
considered.  A general galaxy survey with $r<18.8$ (denoted 
as MAIN1) may be
completed within one  year, while a deeper one with $r<19.8$ 
(denoted as MAIN2) may serve
as the ultimate goal of the LAMOST telescope operation. 
For the QSOs, three magnitude limits
are considered, $g < 20.5$ (QSO1), $g < 21.0$ (QSO2), and  
$g < 21.65$ (QSO3), all between $z=0.4$ and $z=2.1$.
For the LRG survey,
we consider a sample that is similar to the SDSS MegaZ sample \citep{C06}.
Based on the observationally inferred luminosity function and 
HOD parameters,
we estimate the comoving density $n(z)$ and the clustering bias $b$
to determine the effective volume defined in Eq.~(\ref{eqa:Veff}). 
Since the angular density of LAMOST fibers is $200\deg^{-2}$, 
these targets may not be observed in a single run. 
We discuss in more detail the fiber allocation strategy in the Appendices.

\begin{figure}
\begin{center}
\includegraphics[height=0.4\textwidth]{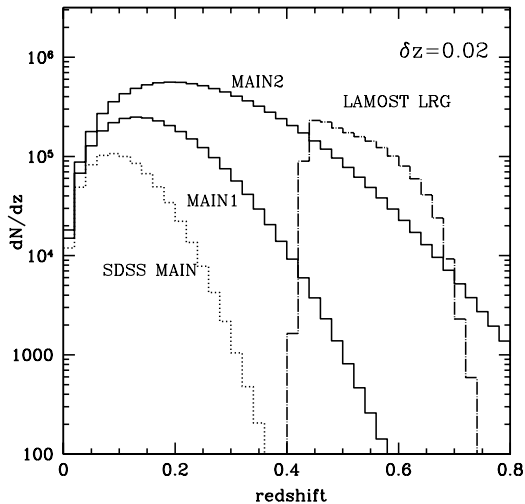}
\caption{\label{fig:dist_galaxy_z} The Redshift distribution of different 
galaxy samples, assuming a sky coverage of $8000\ deg^2$ }
\end{center}
\end{figure}

The redshift distribution of galaxies is shown in 
Fig.~\ref{fig:dist_galaxy_z} (the comoving densities are shown 
in Fig.~\ref{fig:dist_galaxy_den}). We see that 
the redshift distribution of the MAIN1 and MAIN2 samples are broader than the 
SDSS main sample. The minimum redshift of the LAMOST LRGs considered here 
is 0.4. It may be possible to have some more low redshift LRGs, but the present
one is based on a sample which has already been selected (MegaZ). 

\begin{figure}
\includegraphics[height=0.4\textwidth]{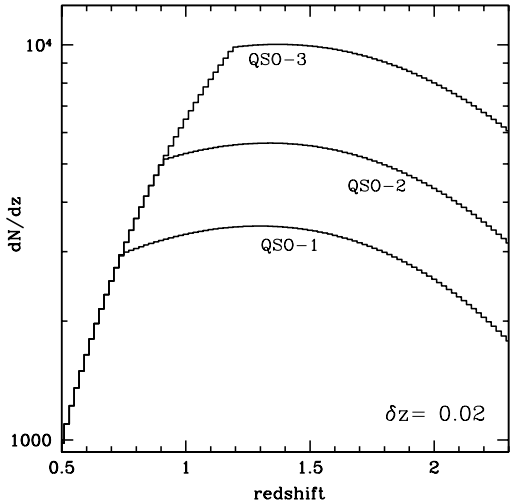}
\caption{\label{fig:dist_QSO_z} The 
redshift distribution of the QSO samples.}
\end{figure}

The redshift distribution of quasars is shown in 
Fig.~\ref{fig:dist_QSO_z} (comoving density is shown in Fig.~\ref{fig:den_qso}).
Compared with the galaxies, the number of quasars is much smaller. 
The drop at low redshift is a combined effect of the existence of
lower absolute magnitude limit (below it an object would be classified as an
AGN rather than a quasar) as well as the decrease of quasar activity.

\subsection{Effective Volume and Distance Scale}

\begin{figure}
\includegraphics[height=0.45\textwidth]{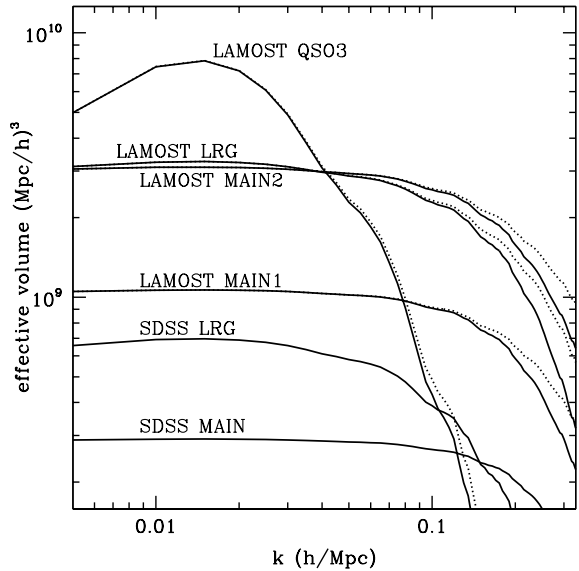}
\caption{\label{fig:Veff} Effective volume of different surveys. 
Solid curves for resolution $R=1000$, dotted curves for $R=2000$.}  
\end{figure}

In Fig.~\ref{fig:Veff} we compare the effective volume of different 
LAMOST survey samples considered in this paper (LAMOST MAIN, LRG and QSO)
and the SDSS samples (SDSS main and LRG\footnote{For the SDSS LRG, 
we assume that the survey 
volume is $0.8(\Gpc/h)^3$ for $8,000~\deg^2$, whereas \cite{SE03}
assumed a volume of $1(\Gpc/h)^3$ for $10000~\deg^2$.  
The density is assumed to be $10^{-4} (h/\Mpc)^3$, and for the bias 
we take the measured value (1.9, as given in \cite{T06}), 
\cite{SE03} originally  
assumed a bias of 2.13. The actual 
survey data may extend to higher redshift, see \cite{E05b,CG08a,CG08b}.}). 
While at small $k$ (large scales)
the effective volumes of the galaxy samples are almost constant, at small 
scales their effective volumes decline. The decline occurs at a scale where
shot noise starts to dominate over the power spectrum (Eq.~\ref{eqa:Veff1}).
So the effective volume of a sparsely distributed sample drops at a scale
larger than that of a high density sample.
The finite spectrograph resolution smears the power spectrum at small
scales and further reduces the signal. We plot the effective volume for both 
$R=1000$ (solid curve) and $R=2000$ (dotted curve) in the figure, and we see
that the decline in effective volume starts at smaller scale 
for the higher resolution setting.

As shown in the figure, the MAIN1 survey
has an effective volume of $1~(\Gpc/h)^3$,
about 3.3 times that of the SDSS main survey on large scales --
it is even 
larger than the SDSS LRG survey as assumed in this paper
(though the actual SDSS LRG survey reaches a 
greater effective volume, see footnote 10).
If the survey of this sample is completed 
within one year as planned, a second round survey could go one magnitude
deeper and obtain the MAIN2 sample, which has the large effective volume of
 $3~ (\Gpc/h)^3$. However, as discussed in Appendix B, collecting this sample
would require much more observing fiber time, since both the number of targets 
and the integration time required for each target increase.

If an LRG sample similar to that of the MegaZ catalog (but extended to 
$8,000 \deg^2$ of the sky) is surveyed, 
the effective volume at large scales is
$3.1 ~(\Gpc/h)^3$, 4 times larger than the SDSS LRG sample. 
While the LAMOST LRG sample has an effective volume comparable to that of 
LAMOST MAIN2 sample, the LRG survey would take much less time.

The shape of the QSO effective volume is different from those of galaxies. 
At large scales, while the effective volume of a galaxy sample becomes flat,
that of the QSO sample increases with $k$. This is because the density of 
QSO is very low and the shot noise always dominates. As a result, the 
effective volume traces the shape of the power spectrum 
(c.f. Eq.~\ref{eqa:Veff1}, with $nP(k) \ll 1$). 
The effective volume of the QSO sample
can reach a peak which is nearly 3 times larger than that of the LRG sample
at $0.02 h/\Mpc$, thanks to the large survey volume. 
However, it decreases quickly towards small scales, and becomes comparable
to the SDSS main sample at  $k\sim 0.1h/\Mpc$. 
Since the decline happens at scales greater than the 
BAO scale, the constraining power of the QSO sample on cosmology is 
relatively weak, although we still expect to detect the BAO signature 
with the QSO sample.

In the above we have discussed the total effective volume of the sample. In 
practice, the survey volume will be divided into several redshift bins. The 
sample density is approximately a constant within each bin, and the 
cosmological distance scale is measured at the center of each bin. 
The measurement errors on the power spectrum and distance scale depend on the 
size of the redshift bins. Larger bins lead to smaller sample variance but
fewer data points on distance measurements, and vice versa. Our tests show 
that the eventual errors on dark energy equation
of state or other cosmological parameters are not sensitive to how we make
the division of the bins, as long as the evolution in density and 
bias within each bin is small. 

In Table~\ref{tab:bin}, we show our division of redshift bins for different 
samples. For the main samples, our bin size is $\Delta z=0.2$, and the MAIN1 
and MAIN2 samples have two and three bins, respectively. 
For the LRGs, the first bin contains the SDSS LRG sample, 
and the other two bins are similar to those in the 
MegaZ catalogue. 
The three QSO samples have different turnover redshifts (determined by the 
apparent magnitude limit as well as the absolute magnitude limit) and they
are divided into slightly different redshift bins.
The bias of each sample is given in Table~\ref{tab:bias}.

\begin{table*}
\caption{\label{tab:bin} Redshift bins of different surveys.}
\begin{tabular}{| c | c | c | c | c | c |c|}
\hline
MAIN1: &(0,~0.2) &(0.2~,0.4)& & & &  \\ 
\cline{1-7}
MAIN2: &(0,~0.2) &(0.2~,0.4)& (0.4,~0.6) & & &  \\ 
\cline{1-7}
LRG: &(0,~0.4) &(0.4~,0.55)& (0.55,~0.7) & & &  \\ 
\cline{1-7}
QSO1:  
 & (0.4,~0.75) &(0.75, ~1.02) & (1.02,~1.29) & (1.29,~1.56) 
& (1.56,~1.83) & (1.83,~2.1) \\
\cline{1-7}
QSO2: 
 & (0.4,~0.66) & (0.66,~0.92) & (0.92,~1.215) & (1.215,~1.51) 
& (1.51,~1.805) & (1.805,~2.1) \\
\cline{1-7}
QSO3:  
& (0.4,~0.67) & (0.67,~0.93) & (0.93,~1.2) &(1.2, ~1.5 ) 
& (1.5,~1.8) & (1.8, ~2.1) \\
\hline
\end{tabular}
\end{table*}

\begin{figure}
\includegraphics[height=0.45\textwidth]{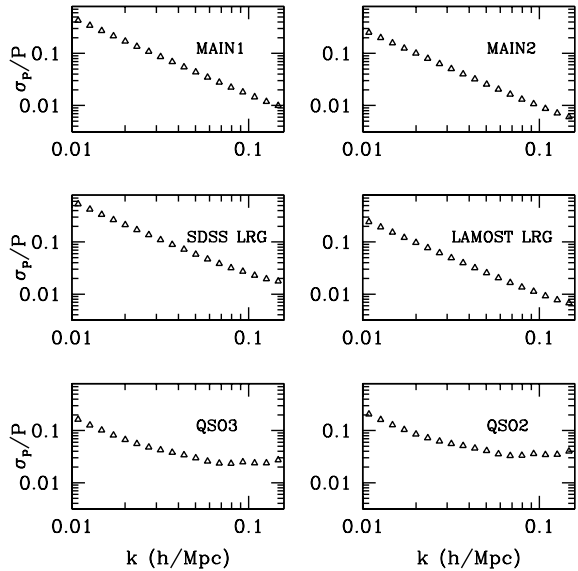}
\caption{\label{fig:errpow}The expected fractional error of the 
galaxy and quasar power spectra. }
\end{figure}

We plot the forecasted matter power spectrum measurement errors 
derived from these samples at different redshifts
in Fig.~\ref{fig:errpow}. For comparison, we also show the expected errors
for the SDSS LRG sample. These errors are calculated assuming that
the whole survey volume is included. For the LAMOST galaxy samples,
the fractional error is at a level of 1\% around $k=0.1 h/\Mpc$.
The errors for the QSO samples are large on small scales due to their low 
space densities, but on large scales ($k\sim 0.01 h/\Mpc$) QSO samples 
yield smaller errors than galaxy sample because of their large survey 
volumes.

\begin{figure}
\includegraphics[height=0.45\textwidth]{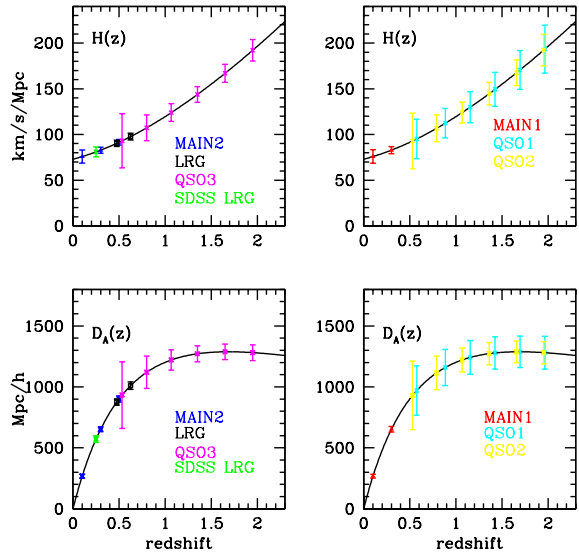}
\caption{\label{fig:DaH} The distance scale $D_A(z)$ and $H (z)$
for the fiducial model and the expected measurement errors. The first LRG 
point is omitted due to coincidence with other data points.
}
\end{figure}

In Fig.~\ref{fig:DaH} we plot the distance scales ($D_A$, $H$) 
and their measurement errors ($\sigma_{D_A},~\sigma_{H}$) as a function of
redshift for the fiducial model. 
The top and bottom panels show the errors for $H(z)$ and $D_A(z)$, 
respectively. 
The measurement errors for the galaxy samples are at the level of a few 
percent. The QSO samples have larger measurement errors, but they can extend 
to higher redshifts.

\subsection{Constraints on Cosmology -- Results from the Fisher Matrix Method}

\begin{figure}
\includegraphics[width=0.45\textwidth]{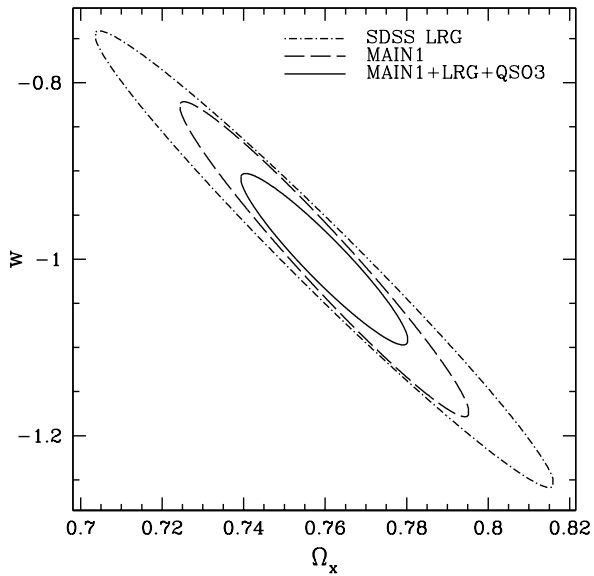}
\caption{\label{fig:DE_constw} Constraints ($1\sigma$) derived from combing 
different data sets for a dark energy model with constant equation of 
state $w$, all assuming the Planck 3-year prior.
}
\end{figure}

\begin{figure}
\includegraphics[width=0.45\textwidth]{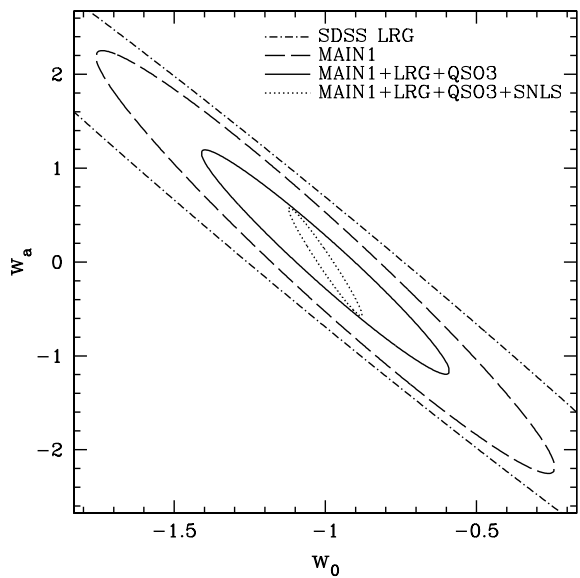}
\caption{\label{fig:DE_comb} Constraints on dark energy parameters $w_0$ 
and $w_a$ derived from combining different data sets, all
assuming the Planck 3-year prior.
}
\end{figure}

Using the projection method described in \S2.1, we derive the error ellipse
on the dark energy parameters after marginalizing over other parameters. In 
Fig.~\ref{fig:DE_constw}, the constraints on a dark energy model of constant 
equation of state $w$ are plotted for the MAIN1 sample and the combination 
of the MAIN1, LRG and QSO3 sample. 
For comparison, we also plot the expected 
error ellipse from the SDSS LRG sample, which has a good constraint on 
the dark energy density parameter $\Omega_X$ and a poor constraint on $w$.
With the LAMOST samples, even for the MAIN1 only sample,
there is a significant improvement in constraining $\Omega_X$ and $w$.
By combining the MAIN1 sample, the LRG 
sample, and the QSO sample, the allowed range in $w$ is halved with respect to
that from the SDSS LRG, reaching a fractional uncertainty of 
about 6.4\%. We did not plot the constraint produced by the MAIN2 sample
in the above figure, but it is almost the same as the combination of 
MAIN1, LRG and QSO3 sample.

\begin{table}
\caption{\label{tab:resolution} $~w_0-w_a$ FoM for different samples
with $R=1000$ and $R=2000$. The fourth column shows the fractional increasing
of FoM by raising the resolution.}
\begin{tabular}{c c c c}
\hline
Sample/FoM &  R=1000   & R=2000  &  +\% \\
\hline
MAIN1   & 5.72  &  5.79 &  1.3\% \\
\hline
MAIN2   & 10.83 & 11.04  &  1.9 \% \\
\hline
LRG  &  7.94 &  8.17  &  2.9\% \\
\hline
QSO1   & 0.554  & 0.638  & 15.2\% \\
\hline
QSO2   & 0.903 &  1.049 &  16.2 \% \\
\hline
QSO3   & 1.425 & 1.673  & 17.4 \% \\
\hline
\end{tabular} 
\end{table}

\begin{figure}
\begin{center}
\begin{tabular}{cc}
\includegraphics[height=0.25\textwidth]{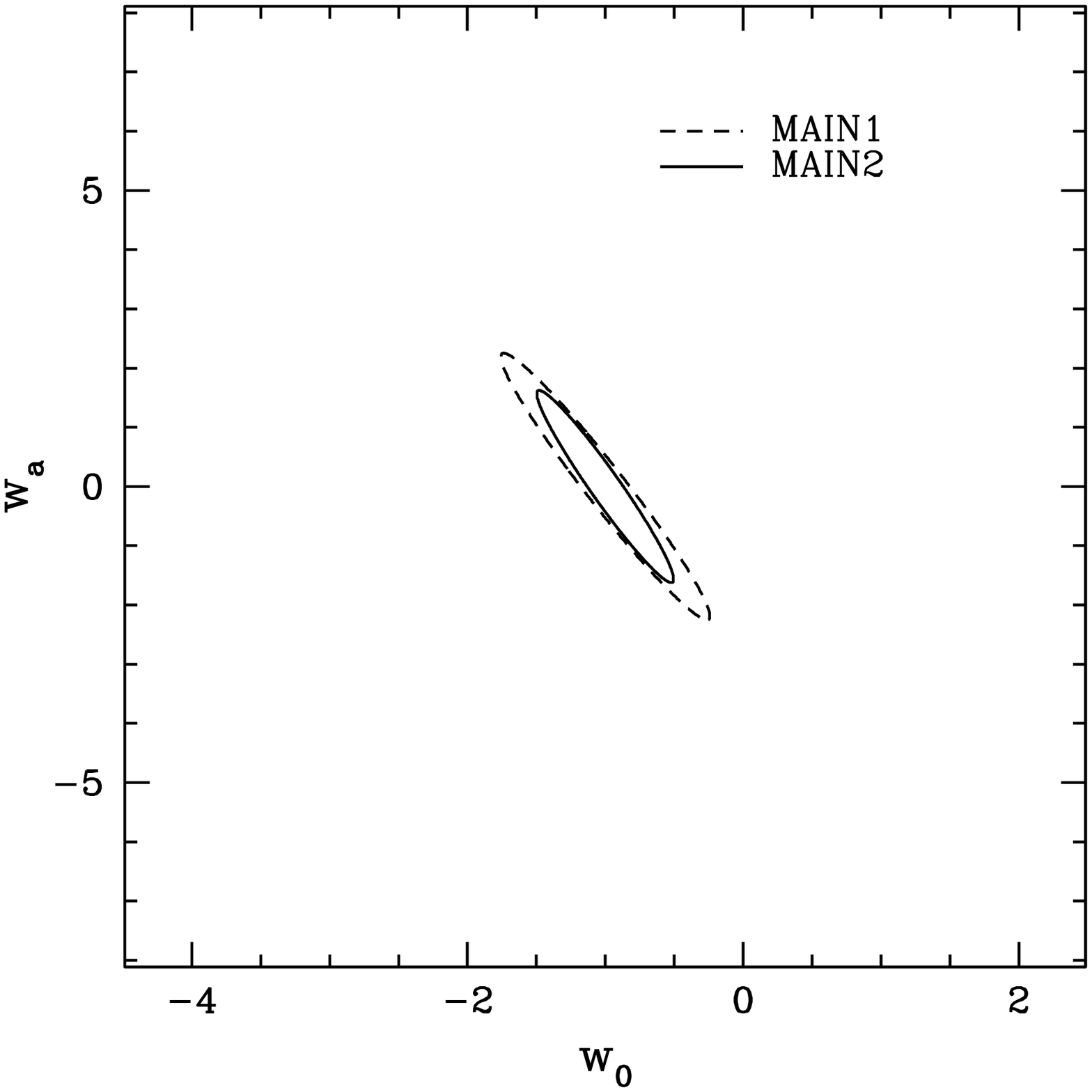}&
\includegraphics[height=0.25\textwidth]{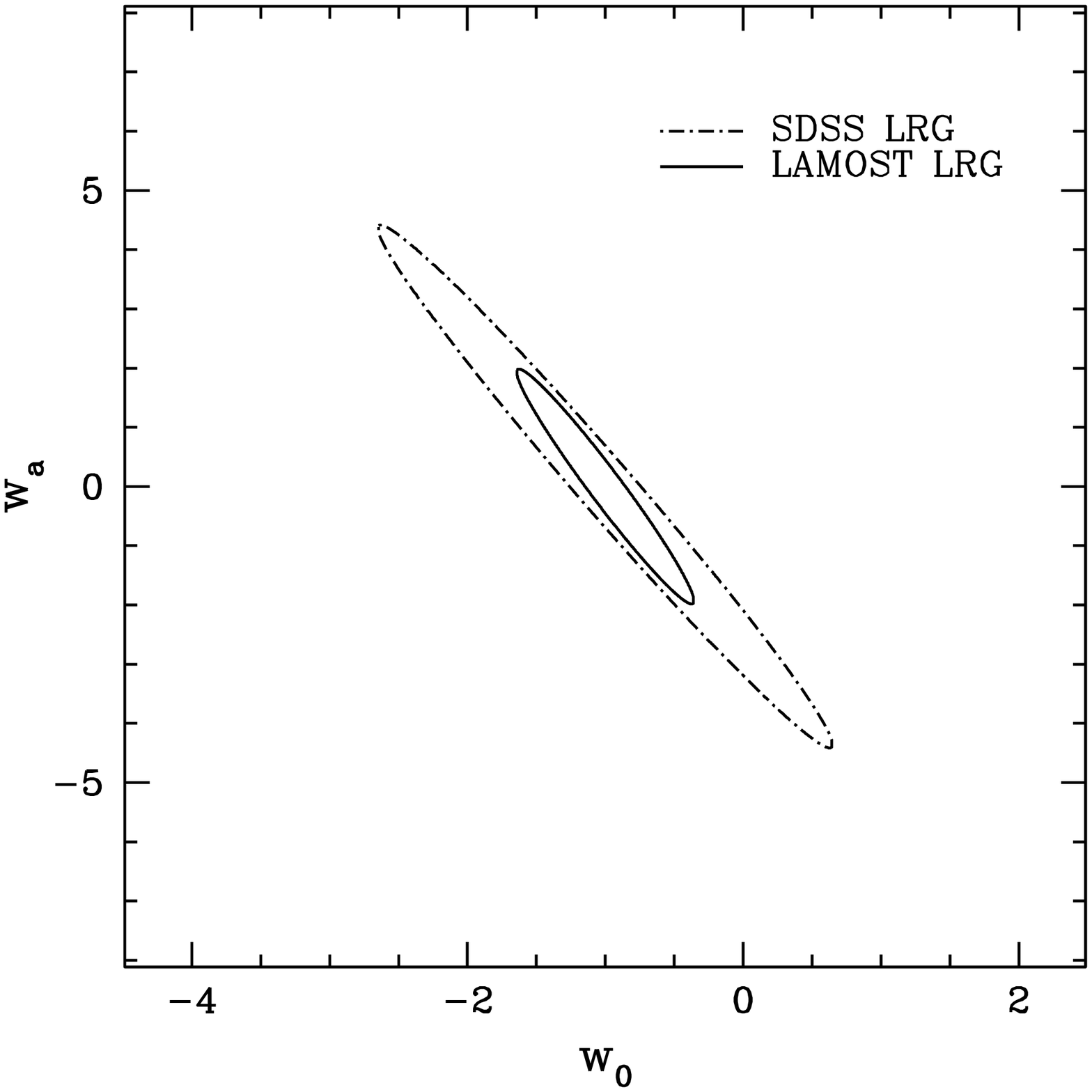} \\
\includegraphics[height=0.25\textwidth]{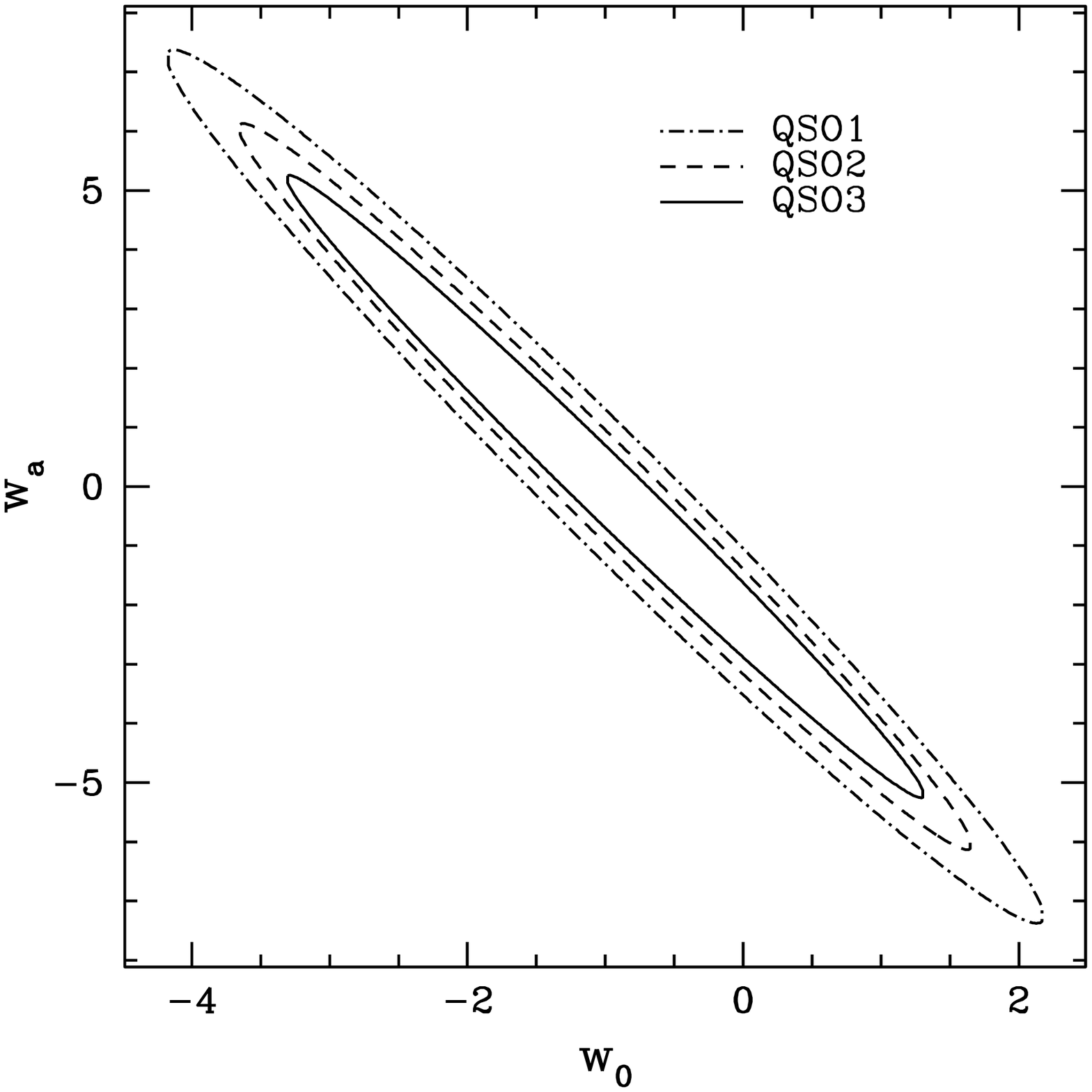}&
\includegraphics[height=0.25\textwidth]{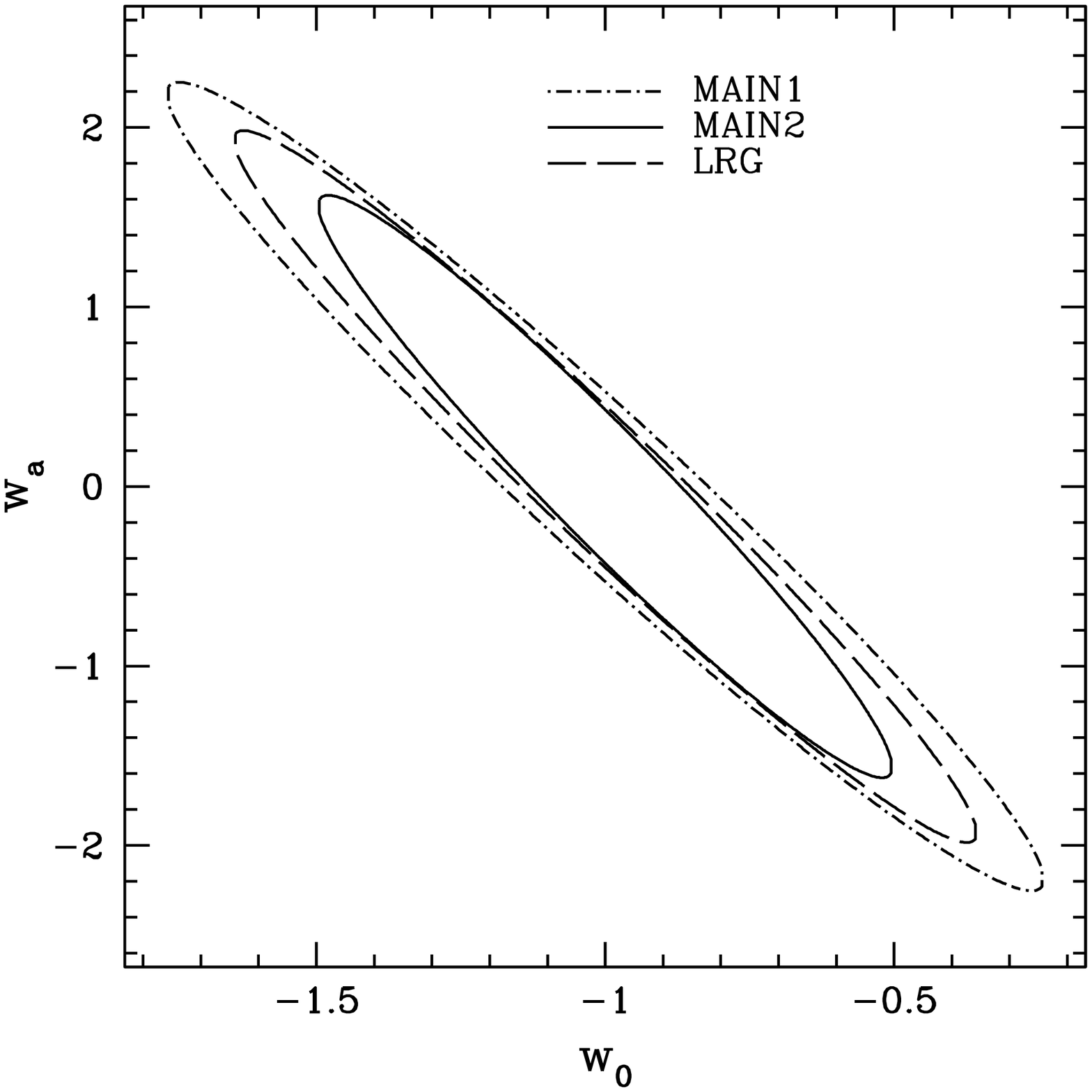}
\end{tabular}
\caption{\label{fig:DE} The $1\sigma$ error ellipse for $w_0$ and $w_a$ for 
different samples: the main samples (top left), LRGs (top right),
QSOs (bottom left), and comparison of main and LRGs (bottom right),
all assuming the Planck 3-year prior.
}
\end{center}
\end{figure}

Next we consider the case of evolving equation of state $w$, 
which is common for dynamical dark energy models. 
In Fig.~\ref{fig:DE_comb}, we show the constraints on dark energy equation of 
state parameters from different combinations of the MAIN, 
the LRG, and the QSO samples. The solid contour shows 
the dark energy constraints that the LAMOST can achieve with all the samples.
There are significant improvements over previous surveys such as the SDSS. 
However, the constraints on evolving $w$ are still weak,
which is in agreement with the assessement of the Dark Energy Task Force (DETF) 
\citep{A06} that  
near term projects could not yet provide high precision measurement on 
$w_0$ and $w_a$. 
By combining the BAO measurements with future Type Ia supernovae results,
the constraints could be greatly tightened, as shown by the dotted contour 
in the plot.

In the above we have considered the constraints derived 
by combining samples of different
LAMOST surveys. How much information does each of the samples provide? 
To address this problem we plot the constraints obtained with each 
of the samples in Fig.~\ref{fig:DE}. We can see that the galaxy samples 
provide stronger constraints than the quasar samples, the constraining power
of the LRG is between the MAIN1 and MAIN2 samples.

We can also define a Figure of Merit (FoM)
\begin{equation}
\FoM=\frac{1}{\sqrt{ \det (\Cov(w_0, w_a) ) }}
\end{equation}
which is proportional to 
the inverse of the contour area\footnote{Our definition follows
that of \cite{W08}. In the Fisher matrix formalism, it is related to the
DETF FoM, which is the inverse of the $2\sigma$ contour area
\citep{A06} as $\FoM_{\rm Wang}=6.17 \pi ~\FoM_{\rm DETF} $.}. 
It is 2.01 for the SDSS LRG sample, 
5.72 for the MAIN1 sample, 7.94 for the LAMOST LRG sample, 
10.83 for the MAIN2 sample. The QSO samples all have very large survey 
volumes, but due to their low densities, their FoMs are small:
$\FoM_{QSO1} = 0.55$, $\FoM_{QSO2} =0.90$, and $\FoM_{QSO3} =  1.43$.
The error ellipses in Fig.~\ref{fig:DE} reflect this trend: the MAIN2 sample
provides the most stringent constraint and the LRG observation could also
achieve a similar FoM with much less fiber time. When combining these samples,
we reach a FoM of $20.88$. Note that since the MAIN2 sample overlaps the 
LRG sample,  we only include the first two bins of MAIN2 in the combination.

The estimate on the FoMs for galaxy samples is not very sensitive to 
the adopted value of the bias factors, as long as the constraining power 
comes mainly from the range of the power spectrum that is not dominated
by shot noise (Eq.~\ref{eqa:Veff1}). 
We test this by adopting observationally inferred
bias factor, instead of that from the HOD method.
If we ignore the difference in the sample definition and use the bias 
factors $b=2.35$ \citep{R07} and $b=1.66$ \citep{R06} for the LRG sample, we 
find that the FoM of the LRG sample changes by 1.8\% and -6.8\%, respectively.
For the QSO3 sample, we find that the change in the FoM is 6.9\% if the 
bias factor measured by \cite{P04} is used in our analysis.

An important question in the design of the surveys is whether to choose
$R=1000$ or $R=2000$. With the higher resolution setting, the redshift could
in principle be measured to higher precision. However, in practice, for a fixed
observation time, a higher signal to noise ratio could be achieved with the 
lower resolution setting. Conversely, if one requires a fixed signal to noise
ratio, more time would be required to complete the survey for the higher 
resolution setting. For some applications, particularly those on the physics
of the galaxies, useful information could be obtained only with the higher 
resolution setting. For dark energy constraint, the difference between 
the two choice is largely quantitative. From Fig.~\ref{fig:Veff}, 
we see that there is almost
no difference on large scales between the two resolution settings. On small
scales, however, the effective volume is larger for the higher 
resolution setting, which shifts the smearing of small scale power 
to smaller scales. From the figure, it seems that for galaxy samples the 
improvement is larger than the QSO samples. However, for the QSO the 
improvement from the higher resolution occurs at the critical scale 
of about $0.05 \sim 0.1 h/\Mpc$, which is exactly where the baryon oscillation
features occur. The figure of merit of dark energy measurement
is given in Table~\ref{tab:resolution} for the two settings. 
For the QSOs, the FoM with $R=2000$ is about $15\%$ bigger than $R=1000$. 
For the galaxies including both the MAIN and LRG samples,
the difference is only a few percent, since the improvement is made
largely in the non-linear regime. Therefore, from the perspective of dark 
energy constraints, $R=1000$ is sufficient and preferred.

\subsection{Constraints on Cosmology -- Results from the MCMC Method}

\begin{figure}
\includegraphics[width=0.45\textwidth]{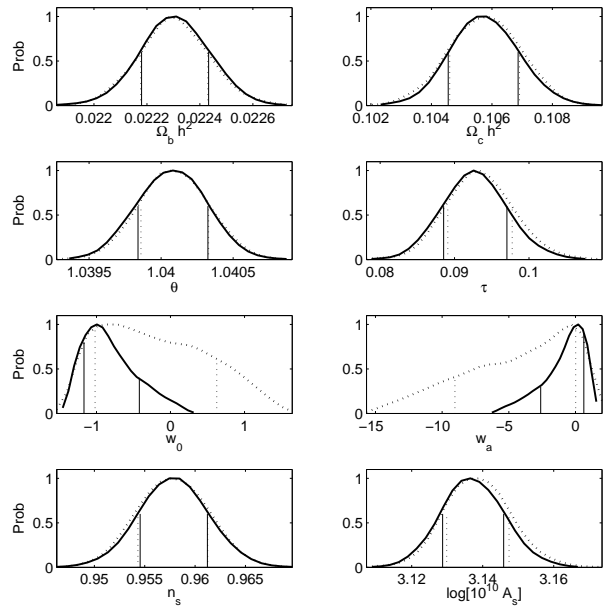}
\caption{\label{fig:1dmcmc}The probability distribution function for some
cosmological parameters constrained by combining the CMB and BAO data.
The solid and dashed curves correspond LAMOST LRG (MegaZ) sample and the
SDSS LRG sample. The 68.3\% distribution is marked with two verical lines.}
\end{figure}

\begin{figure}
\includegraphics[width=0.45\textwidth]{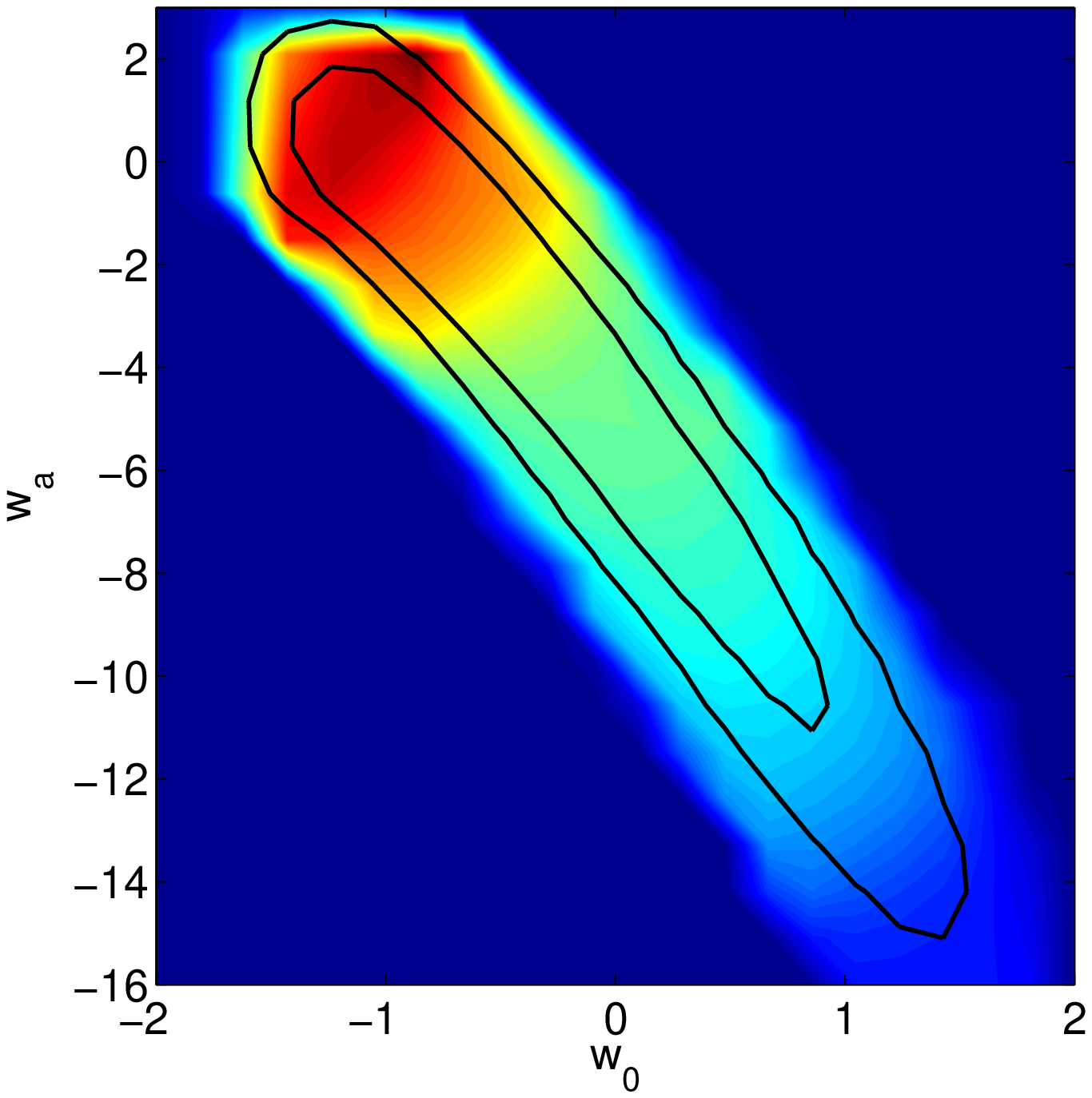}
\includegraphics[width=0.45\textwidth]{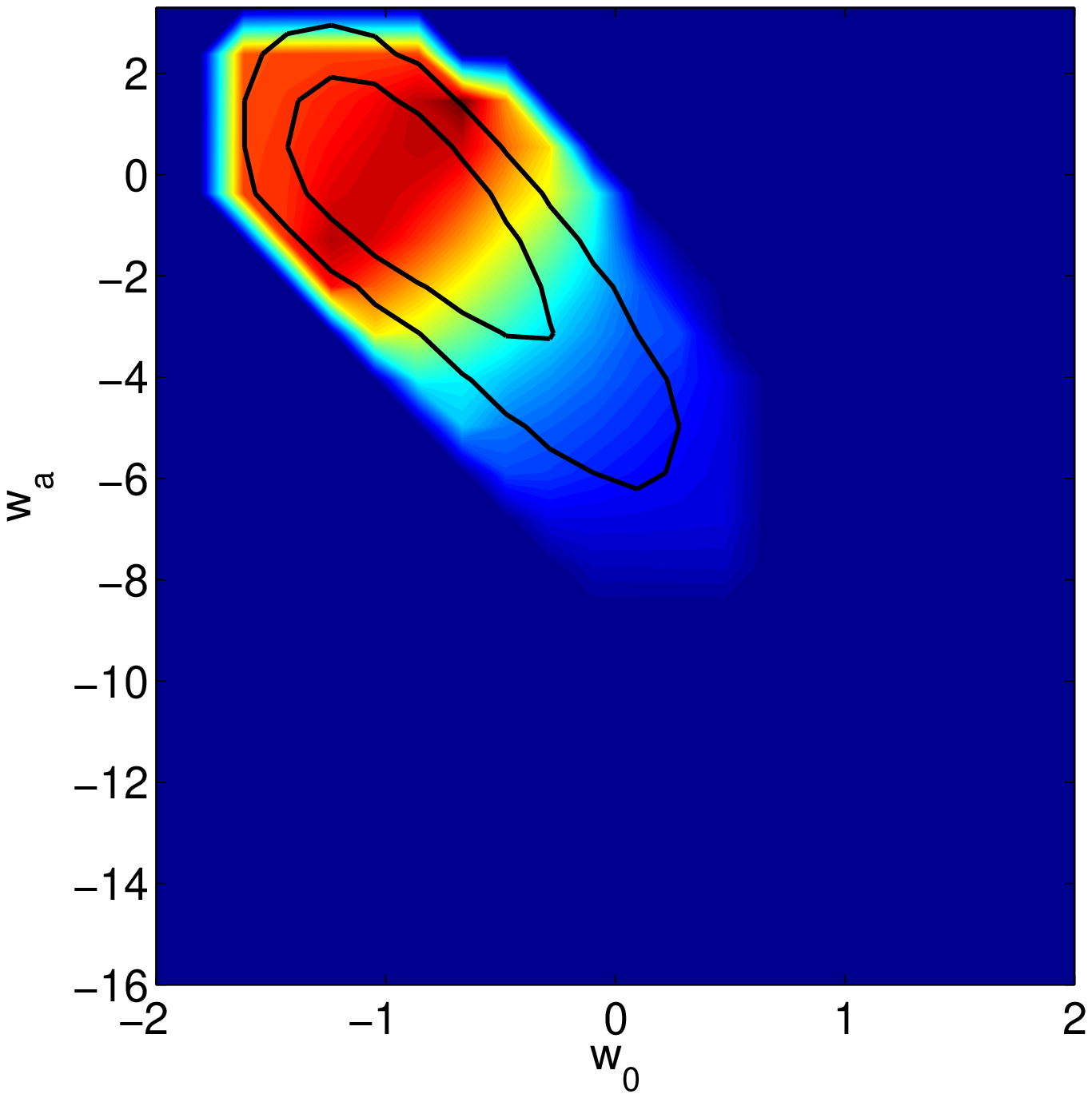}\\
\caption{\label{fig:mcmc} The expected constraints on dark energy 
equation of state parameters by combining the CMB data and the BAO measurements
with the LRG sample. The BAO measurements in the top and bottom panels 
are assumed to come from the SDSS-II LRG and the 
LAMOST LRG samples, respectively.
Contours in each panel correspond to 68.3\% and 95\% confidence levels,
and the shading shows the mean likelihoods. 
}
\end{figure}

We have also performed Monte Carlo simulations for a few cases.
In particular we consider the SDSS LRG, which represents our currently 
available data, and the LAMOST LRG, which
represents the precision achievable after two or three years of LAMOST 
survey. We derive constraints for 
two cases: (1) only the CMB and BAO data are used; (2) CMB, BAO and SNIa data
are used. In the first case, only the ``clean'' methods
(i.e., the CMB and BAO) that are based on well understood physics
are used. 
In the second case, 
we include a sample of 1200 supernovae expected from SNLS.

In Fig.~\ref{fig:1dmcmc} we show the probability distribution function (PDF) 
for each of the interested cosmological parameters obtained
from the first case after marginalizing over the rest of the parameters.  
The solid (dashed) curves are from CMB combined with the LAMOST LRG 
(SDSS-II LRG) sample.
Except for those of $w_0$ and $w_a$, the PDF is well approximated
by a Gaussian distribution. There is no significant 
improvment in the constraints on  
cosmological parameters other than $w_0$ and $w_a$ by the addition of the 
LAMOST data, 
as the constraints from the CMB data are already very stringent.
Compared with the SDSS-II LRG sample, the LAMOST LRG sample helps to 
significantly tighten the constraints on the dark energy parameters.

\begin{figure}
\includegraphics[width=0.43\textwidth]{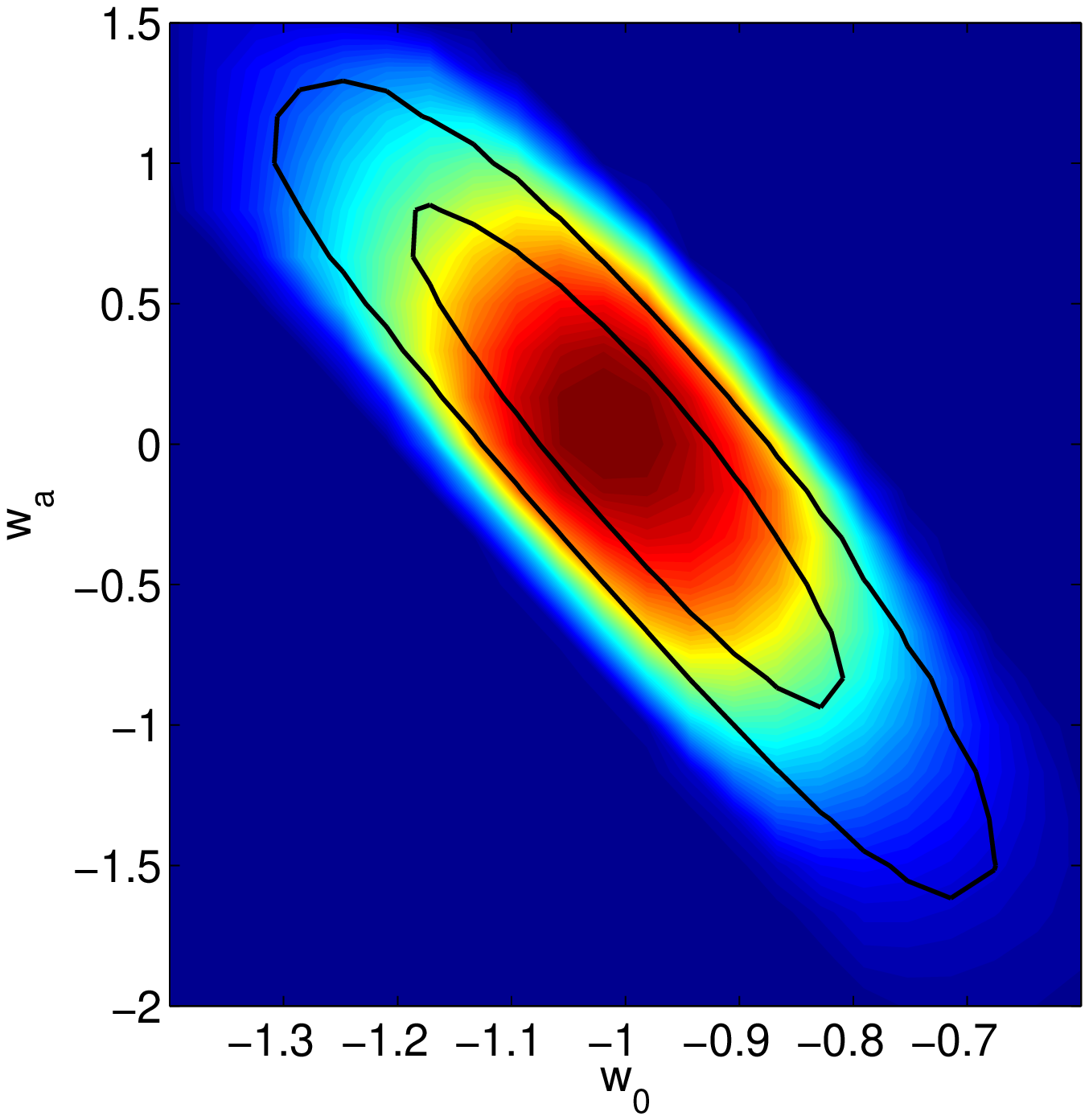}
\includegraphics[width=0.43\textwidth]{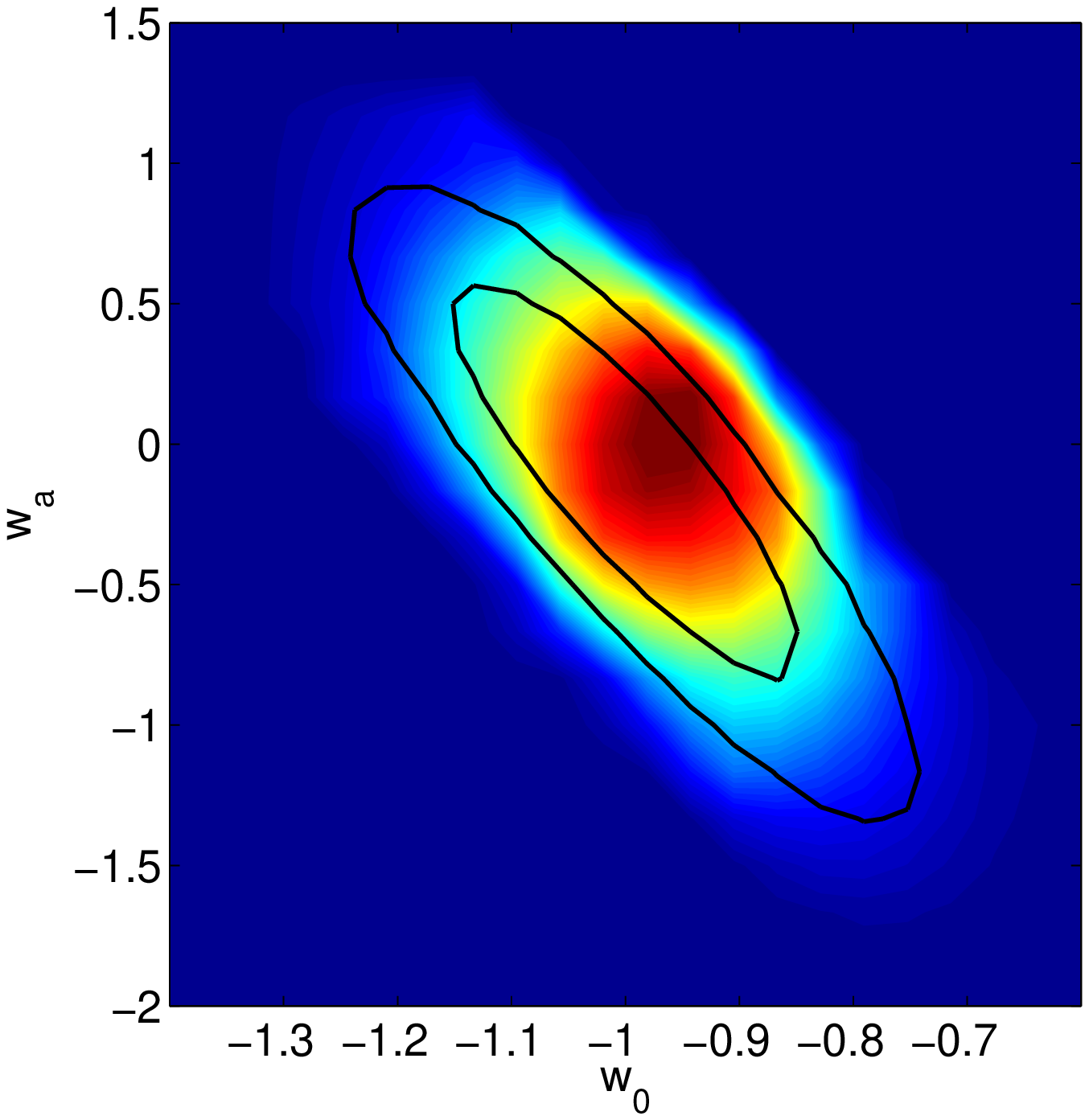}\\
\caption{\label{fig:mcmcsn} Top: Constraints on $w_0$ and $w_a$
by combining the CMB (Planck) and the SNIa (SNLS) data. Bottom:
Constraints on $w_0$ and $w_a$ by combining the CMB (Planck),
SN Ia (SNLS) and also the BAO (LAMOST LRG) data.
Contours correspond to 68.3\% and 95\% confidence levels,
and the shading shows the mean likelihoods. }
\end{figure}

In Fig.~\ref{fig:mcmc} we show the constraints in the $w_0$--$w_a$ plane.
The contours correspond to 68.3\% and 95\% confidence levels,
and the shading is the mean likelihoods of samples \citep{LB02}.
With the LAMOST LRGs, the contours are 
much smaller than those with the SDSS-II LRGs. 
For our choice of non-linear cut-off scale, the size of the contour generally 
agrees with the Fisher matrix estimation, but not exactly. In particular, 
for SDSS-II LRGs, the long tail towards the lower right corner reflects the 
non-Gaussinity of the likelihood distribution.  With only the 
CMB and BAO data measured by LAMOST, the constraint on dark energy 
is not very stringent, and many different dark energy model would have 
easily passed the test.

To improve the constraints on dark energy parameters, we
can bring in the SNIa data. To separate the contributions from
the SNIa and BAO data, we first show the constraints with CMB and
SNIa data in the top panel of Fig.~\ref{fig:mcmcsn}. Comparing with the top
panel in Fig.~\ref{fig:mcmc} (CMB+BAO), we see that the SNIa data provide
significant complementary information. The bottom panel of 
Fig.~\ref{fig:mcmcsn}
shows the constraints from all the three data sets (CMB+SNIa+BAO).
Compared with CMB+SNIa, adding BAO data from LAMOST LRG further
tightens the constraints, increasing the FoM from 33.07 to 42.83. 
Here, the FoM is defined as $6.17 \pi$ over the contour area of 95\% 
confidence level.
Such constraints may be able to distinguish some dynamical dark energy
models.

\section{Conclusion}

In this paper we forecast the measurement precision of dark energy 
equation of state from the BAO measurement expected from LAMOST surveys. 
We investigate three types of sample 
targets: magnitude limited surveys of all types of 
galaxies (main survey), a survey of color-selected luminous red galaxies 
(LRG), and surveys of quasars (QSO). For the main survey and the 
QSO survey, we consider a few different magnitude limits, all deeper than the 
SDSS spectroscopic survey. For the LRGs, we have used the MegaZ sample 
as our reference sample. The required fiber hours to 
complete each survey are also estimated (Table\ref{tab:summary}).

For each sample, we predict the redshift distribution,
and use the halo model to estimate their clustering 
bias. We then calculate the effective
volume of each survey as a function of wavenumber $k$, 
the statistical errors in the measured matter power spectrum, and the 
resulting errors in distance scale measurements with the BAO used as 
standard rulers.
The constraints on the dark energy equation of 
state parameters are derived using two methods, the Fisher matrix method
and the MCMC method.

Our results yield useful insights on how to design the surveys. 
We find that the MAIN1 sample, which is one magnitude deeper than the 
SDSS main sample has an effective volume about three and half 
times larger than the SDSS main sample. Similarly, with the LAMOST LRGs, 
an effective volume of about four times of the SDSS LRG can be 
obtained. This effective volume is almost as large as the much more 
time-consuming MAIN2 sample (two magnitudes deeper than the SDSS main sample). 
The QSO samples can also yield high 
effective volume at large scales, but it declines rapidly at small scales.
The improvement on the measurement of dark energy equation of state 
depends not only on the large scale effective volume, but also on the 
effective volume at relatively 
small scales. Our investigation shows that although 
the effective volume of the LAMOST LRG sample is comparable to the 
MAIN2 at large scales, the latter's
figure of merit for dark energy measurement is greater. 
The QSO sample, on the other hand, 
generally has lower figure of merit. However, the number of targets and
required observing fiber hours of the QSO samples are also much smaller
than the galaxy samples, so it should not be regarded as a competing project
of the galaxy surveys. The efficient design for using LAMOST survey 
to constrain dark energy parameters is to have a MAIN1 survey, an LRG survey
supplemented by a QSO survey.

We also studied the 
impact of spectral resolution. The figure of merit improves by only about 
1\%-3\% for the galaxy samples 
when one switches the spectral resolution from 1000 to 2000. For quasars
the figure of merit improvement 
is about 15\%. 
From the point of view of dark energy constraints, 
it is unnecessary 
to achieve a high spectral resolution,
although high resolution spectra are useful for other scientific 
objectives.  

Finally, we also used Monte Carlo simulations to make the same forecast
to test the accracy of the Fisher matrix result.
The figure of merits of the 
two estimate roughly agree, although the contours obtained from the Monte Carlo
method are not exactly ellipses, but rather asymmetric.

Overall, the LAMOST surveys could achieve an figure of merit (as defined by 
\cite{W08}) of 5-10 from each individual galaxy samples. For a constant $w$ 
dark energy model, the expected precision on $w$ is about 6.9\%. 
For time-dependent $w$ models, the LAMOST survey would also help to 
significantly improve the constraints on the equation-of-state parameters,
although the expected errors on $w_0$ and $w_a$ are not tight enough to
pin down the nature of dark energy.

\section*{acknowledgments}
We thank Yuan Xue and Huoming Shi for providing us the
required LAMOST exposure time for different magnitudes; and to 
A. A. Collister and his colleagues, for providing us the MegaZ catalogue. 
We are also grateful to Yaoquan Chu, Xinmin Zhang, Hong Wu, Haijun Tian, 
Haotong Zhang, Yipeng Jing, Xianzhong Zheng, 
Hu Zhan, Junqing Xia, Hong Li, Zuhui Fan,  
H. J. Seo, Daniel Eisenstein, Chris Blake, Jonathan Pritchard and Michael 
Strauss for
helpful discussions.  Our MCMC computation are performed on the 
Supercomputing Center of the Chinese Academy of Sciences and the Shanghai
Supercomputing Center. This work is supported by
the National Science Foundation of China under the Distinguished Young
Scholar Grant 10525314, the Key Project Grant 10533010, by the
Chinese Academy of Sciences under the grant KJCX3-SYW-N2, and by the 
Ministry of Science and Technology national basic science
Program (Project 973) under grant No. 2007CB815401.  Z. Z. gratefully 
acknowledges support from the Institute for Advanced Study through a 
John Bahcall Fellowship.

\appendix 


\section{The LAMOST telescope}
The LAMOST telescope (see e.g., \citealt{C98,R08}) is a 4 meter
 telescope\footnote{The main reflector of the LAMOST has a diameter of 
6 meters. The mirrors of LAMOST are hexagonal, the effective aperture 
for the whole telescope varies between 3.8 m to 4.5 m depending on  
the position of the pixel on the focal plane and the altitude 
of the target. Calibration for fibers on different part of the 
focal plane is provided by observing standard stars at 
real time.}, its
design features an fixed horizontal meridian reflecting Schmidt 
configuration, with an active Schimidt correcting plate to achieve 
both wide field of view ($5\degr$ diameter, or $20 \deg^2$) 
and large aperture. The focal plane accommodates 4000
optical fibers for spectrographic observations. 
The optical axis is fixed in the meridian plane, taking measurements
of targets as they pass through the meridian (typically a target 
can be tracked for about 2 hours.) The telescope is located at 
the Xinglong station ($117\degr 34\arcmin 30\arcsec$ East,
$40\degr 23\arcmin 36\arcsec$ North) of
the National Astronomical Observatory of China, and can
observe the part of sky with declination $-10\degr <\delta < 90\degr $. 
The construction of the LAMOST telescope was completed by August 2008 and
it is undergoing engineering tests and calibrations. The surveys are 
expected to begin in 2010/2011 and last for at least 5 years.

The LAMOST is to be equipped with both low resolution spectrographs and medium
resolution spectrographs \citep{ZX00,ZW04}, 
the latter is used mainly for bright targets such
as stars. For extragalactic observations using the low resolution 
spectrographs, spectra can be taken with two resolution settings: 
(1) using the full sized fiber, the resolution is 1000 
(2) using half sized fiber, the resolution is 2000. 
The main features of the telescope is summarized in Table \ref{tab:LAMOST}.

The LAMOST has a parallely controlled fiber positioning system with 
4000 optical fibers\citep{X03,ZX04}, 
each individually driven by two micro-stepping
motors and can move within a circle of 30 mm diameter 
(corresponding to $340 \arcmin$ on the sky) to aim at a target
within the circle. The distance between two adjacent circle axes are
26 mm, slightly smaller than the diameter of the circle, so that adjacent
circles overlap. This allows full coverage of the focal plane. Due to 
fiber collision,  targets within 55 arcsec can not be observed 
at the same time. The average angular density of the fibers 
is about 200 per square degree.
The whole assembly is controlled by computer and 
can be adjusted in real time. Acquiring new target and 
adjust the fibers takes a few minutes typically.

As the LAMOST telescope is dedicated to spectroscopy and does not have imaging
capability, photometric catalogue compiled from other surveys
would be used to select targets. The LAMOST will carry out several different
survey projects simultaneously. The input catalogue of these different 
surveys will be combined (with different assigned priority) 
to produce the general input catalogue of LAMOST. 
A software system \citep{PX04} would 
then select targets of observation from this catalogue, taken into account 
the priority of the target, the observing time and target area, the 
lunar phase and weather condition etc, this allows more efficient use
of the fibers and observing time. During each run, the selected targets are 
a mixture of different survey projects, so the appropriate
measure of observing resource is the fiber time, i.e. the number of 
fibers allocated multiplied by the observing time.

\begin{table}
\caption{\label{tab:LAMOST}Main Characteristics of LAMOST}
\begin{tabular}{l r}
\hline
Average aperture  & 4 m\\
Field of view   & $5\,^{\degr}$\\
Focal plane     &  $\phi 1.75$m \\
Focal length    &  20 m\\
Number of fibers & 4000\\
Spectral ranges  & 370-900nm\\
Spectral resolution &  1000 or 2000 \\
Minimal distance of adjacent targets: & $55 \arcmin$\\
Observable sky   & $-10\,^{\degr}< \delta< +90\,^{\circ} $\\
\hline
\end{tabular}
\end{table}

The collecting area of LAMOST is about 2.2 times that of the SDSS 
telescope, which has an aperture of 2.7m. We would expect that for 
comparable exposure time, target of one magnitude deeper could be observed
 if the system efficiency and sky background are the same. 
The system effeciency, in turn, also depends on the positioning 
error of the fiber. The signal to noise ratio of
an observation with different conditions and parameters has been 
estimated using ray tracing method, and a  
calculator has been developed for estimation \citep{XS08}.
For the present paper, we assume an average fiber 
positioning error of $0.5\arcsec$, and a sky background of 
20.5 in r band 
\footnote{Recent observations at Xinglong station show a sky background 
between the 20.5 and 21.0 during moonless nights (H. Wu, 
pravite communication).}. 
We also assume that an average
signal to noise ratio of 4 per spectral resolution 
pixel is achieved for both galaxies and point sources. 
This is about the minimum for redshift 
measurement. Surveys with low signal to noise ratio allows fast survey 
of a large number of targets, however, in the real survey, a higher 
signal to noise ratio may be required for other scientific purposes. 

\begin{figure}
\includegraphics[height=0.45\textwidth]{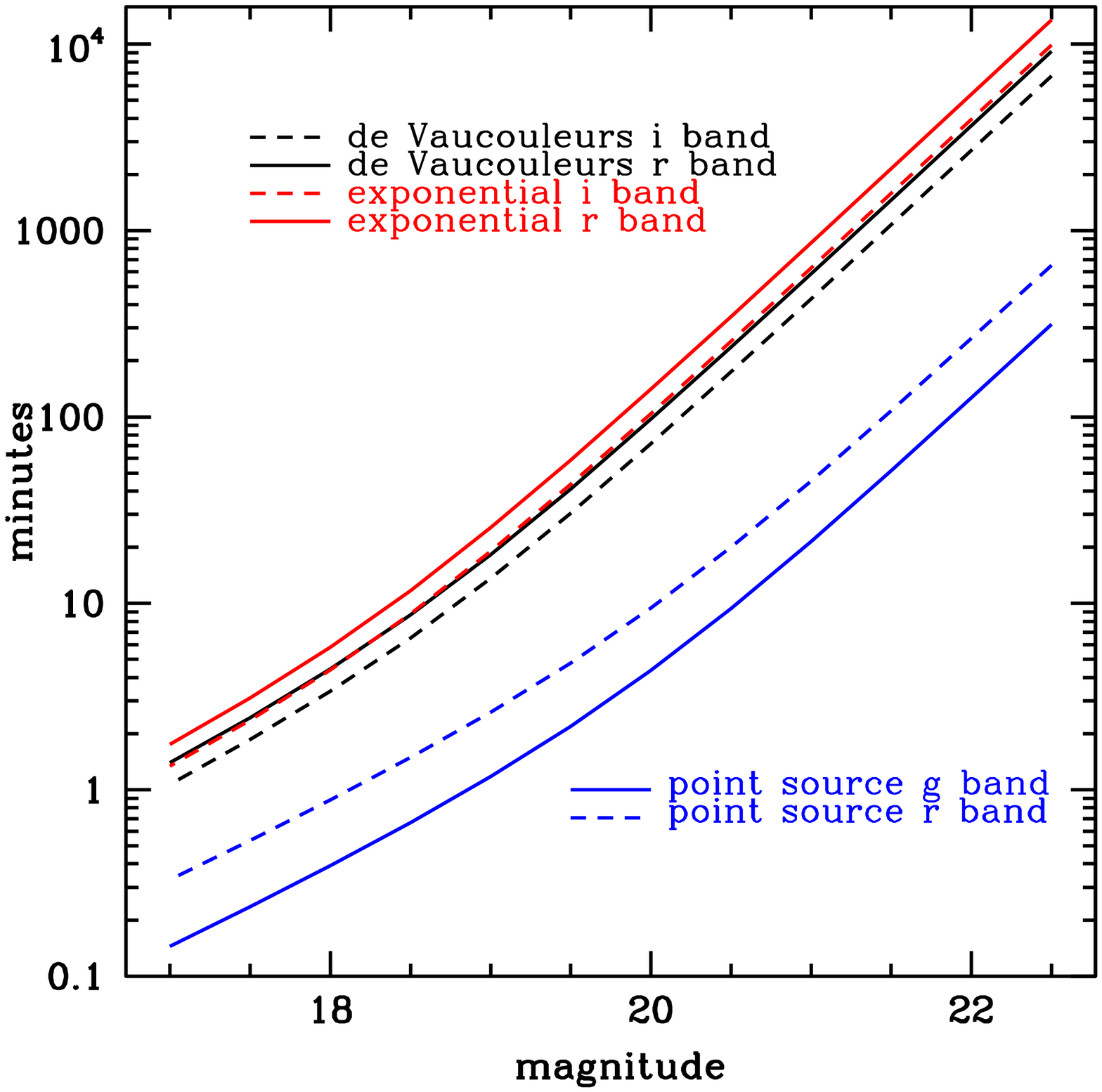}
\caption{\label{fig:expt} Expected exposure time of different bands and
         model profiles, assuming spectral resolution 1000, 
S/N=4 per spectral pixel, a sky background
of 20.5 in the respective band (r or i), 
fiber positioning error $0.5\arcsec$, $r_e=3\arcsec$ for exponential
profile and $r_e=4\arcsec$ for de Vaucouleurs profile.}
\end{figure}

The required exposure time for achieving the desirable signal to noise ratio 
depends also on the size of the object (characterized by the half-light
radii of the image $r_{50}$ for extended source), 
the seeing of the sky, and the point spread function of the optical system. 
We characterize the angular size of the target by a parameter 
$r_e$. For the de Vaucouleurs profile which represents elliptical galaxies,
 $$I_{\rm deV}(r) = I_0 \exp{ [-7.67(r/r_e)^{1/4} ] },$$
 and for the exponential profile which represents spiral galaxies,
$$I_{\exp}(r) = I_0 \exp{ [-1.68 r/r_e ] }. $$
Analysis of the half-light radii ($r_{50}$) of SDSS galaxies show 
that \citep{B01} about 80\% galaxies have $r_{50,petro} \le 3\arcsec$. 
For the exponential profile
$r_{50,petro} = r_{50,true} = r_e$, and for the de Vaucouleurs profile, 
$r_{50,petro} \approx 0.7 r_{50,true} = 0.7 r_e$. Thus, for a simple estimate
of the size of the targets, we assume that the average target 
has $r_e=3\arcsec$ and $4\arcsec$ for exponential profile
and de Vaucouleurs respectively.

In Fig.~\ref{fig:expt}, we plot the exposure time required for 
targets of different sizes, bands and model profiles. 
As can be seen from the figure,
for a given S/N, the integration time is proportional to 
$f^{-1}$ for bright sources, where $f$ is the flux, and to $f^{-2}$
for the faint sources where the sky background noise dominates.
Compared with elliptical galaxies, the
spiral typically requires more time for observation to reach the 
same magnitude, this comprises about 25\% - 50\% of our sample.
Observing large samples of targets fainter than $r=21$ would be impractical. 

The spectral resolution of the low-resolution grating is 1000 when the full
sized fiber is used, 2000 when half sized fiber is used. When observing a
galaxy or quasar with multiple lines, the redshift could be 
measured with accuracy better than the grating resolution if multiple 
spectral lines are present and the signal to noise ratio is high. However, 
the number of usable lines may be limited for many galaxies, and to save
observing time the signal to noise ratio is perhaps not very high 
in large scale observations, so we assume 
the redshift measurement precision to be 1/1000 or 1/2000.

\section{Design of the extragalactic surveys}

Since the LAMOST does not have imaging capability by itself, the targets
of the spectroscopic survey must be selected from photometric catalogues
compiled from observational data obtained by 
other instruments. At present, the SDSS photometric catalogue is the 
largest of such catalogue, which covers about $8000 \deg^2$ of northern sky 
up to $r=23$, and 
only a small fraction of the objects on this catalogue has been
targeted by the SDSS spectroscopic surveys. One may also consider to 
conduct a photometric survey for a part of the south galactic cap 
to supplement this. Regions of low 
galactic latitude are of less interest for extragalactic surveys due to the 
higher extinction.
In the future more targets could be obtained from photometric surveys which
are being planned, e.g., the Pan-Starrs surveys. However, as 
discussed in Appendix A, it is unlikely for LAMOST to observe a large number of
very faint objects within the limited observation time available.

In the present work, we shall consider three surveys: 
(1) a magnitude limited general 
survey of galaxies of all types which we shall call the main survey;
(2) an LRG survey; (3) a magnitude limited low redshift $(z<2.1)$ 
quasar survey.
In Table ~\ref{tab:summary} we list 
a summary of these surveys, including the surface density, total number of 
targets, and required fiber hours.

\subsection{The main survey}
In the original feasibility study of the LAMOST, 
a typical number quoted is B=20.5
for an exposure time of 1 hour, this is comparable to the present
estimate. The SDSS team has chosen their main sample using an r-band
limit, as the r-band luminosity is a better tracer for mass. 
Here we assume the survey limit is given in r-band.
The magnitude limit we quote here is
just a strawman value and can be changed for the real survey.
It is reasonable to consider a sample which is
one or two magnitudes deeper than the SDSS main sample. The SDSS main sample
has a magnitude limit of r=17.77, the exposure time is typically 
about 45 minutes \citep{ST03}.
We consider two samples: a magnitude limited sample 
which is one magnitude deeper than the SDSS main sample, i.e.
$r<18.8$ (hereafter MAIN1), and also a sample which is 2 magnitudes deeper 
than SDSS main, i.e. $r<19.8$ (hereafter MAIN2).

Given the above magnitude limit, we may estimate the angular density
of targets using galaxies number count \citep{Y01} or the 
observed luminosity function \citep{B03}.
\cite{Y01} have shown that the observed number of galaxies 
$N(m)$ can be well fitted by assuming the Euclidean geometry for 
three-dimensional space, 
which gives the number of galaxies brighter than $m$ 
\begin{equation}
  N(m) \propto  10^{ 0.6 m }
\end{equation}
Alternatively, 
the luminosity function of SDSS galaxies was given in \citet{B03}
\begin{equation}
\Phi(M,z) = 0.4 \ln{10} \times \Phi^{\star} 10^{0.4(\alpha +1) (M^{\star}- M) }
     \exp(-10^{0.4( M^{\star} - M)  }  )
\end{equation}
and the redshift evolution of the luminosity is modelled by 
\begin{eqnarray}
 \Phi^{\star}(z)&=&\Phi^{\star}(z_0) 10^{ 0.4P(z-z_0) }  \\
  M^{\star}(z) &=  & M^{\star}(z_0) - Q(z-z_0)
\end{eqnarray}
The absolute magnitude $M$ is related to the apparent magnitude
$m$ and redshift z as 
\begin{equation}
M - 5\log_{10} h = m- DM(z) - K(z)
\end{equation}
where $DM(z)$ is the distance modulus and $K(z)$ the K-correction. 
The K-correction depends on the SED of the galaxy,
we have adopted a simple approximation
of $K(z)\approx -0.21+1.1z$, which is approximately the median K-correction
for the 12 types of galaxy SEDs given in \cite{B03}. This 
may not be very accurate for 
individual galaxies, but should suffice for the purpose of estimating 
the angular density.

\begin{figure}
\includegraphics[height=0.45\textwidth]{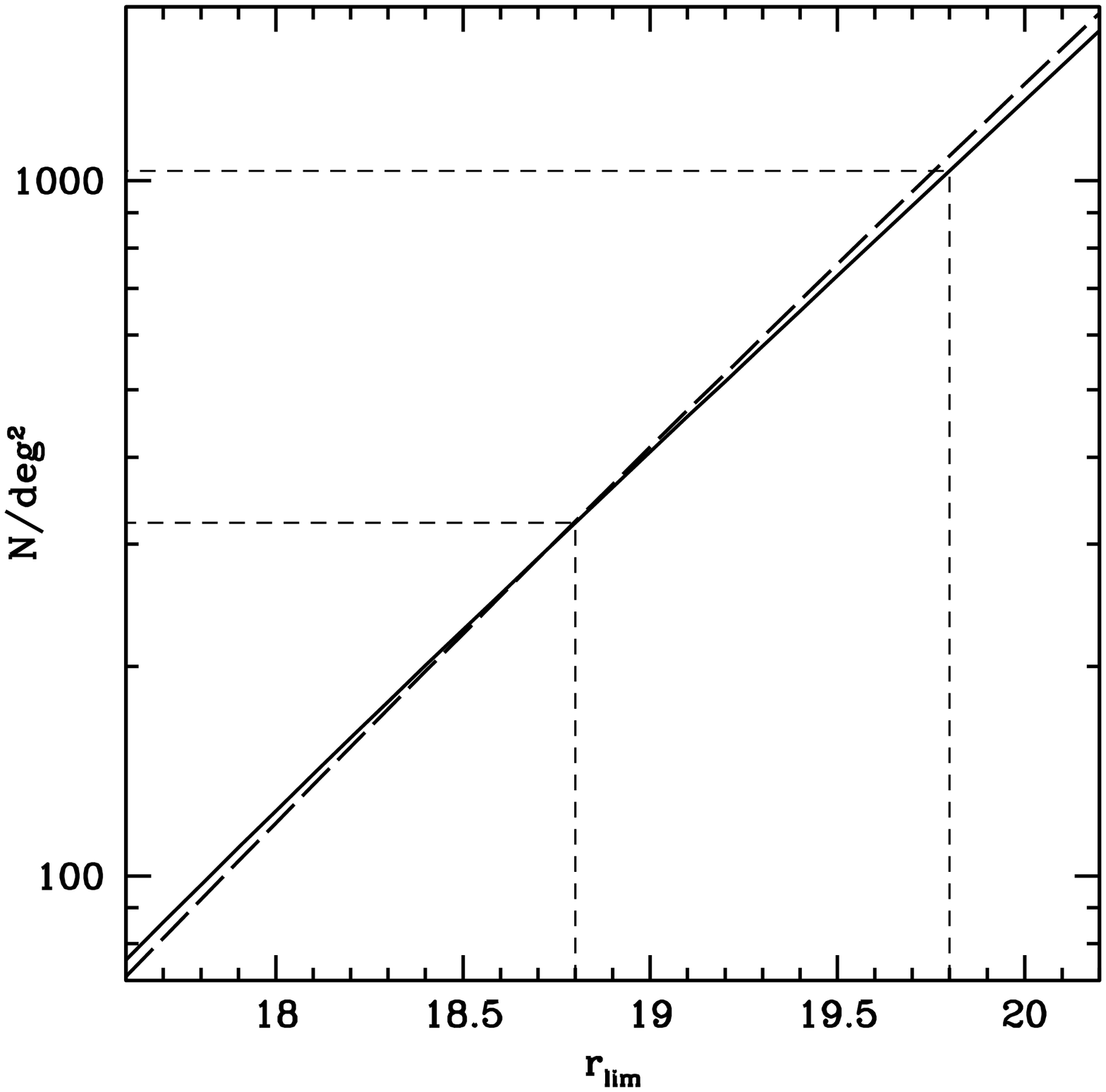}
\caption{\label{fig:angdens_main} The angular density of objects as a function 
of limiting magnitude in the general galaxy survey.
The solid line corresponds to the (apparent) magnitude integration of
the galaxies number count \citep{Y01} at SDSS commission stage. 
The dashed line is the result of (absolute) magnitude-spatial integration
of the luminosity function \citep{B03}, 
by using the best-fitted $r$ band evolution parameter 
$Q=1.62, ~ P=0.18$ and a simple K correction model.}
\end{figure}

We show in Fig.~\ref{fig:angdens_main} the integrated angular density as a 
function of the magnitude limit in $r$ band.
The solid curve is obtained by using 
the galaxies number count given in \citet{Y01}, 
and the dashed curve is obtained by using the luminosity function 
of \citet{B03}. These data have to be extrapolated to faint end or high
redshift to obtain the distribution function. We use the best-fitted $r$ band 
evolution parameter $Q=1.62, ~ P=0.18$ given in \cite{B03}.  
In this paper, we have used the number count directly for 
the fibre time estimation, but for cosmological constraints we used the 
luminosity function. There is only slight difference between these two.

For the MAIN1 ($r=18.8$) sample, the angular density is $330 \deg^{-2}$,
about 1.65 times the average angular density of the fibers. Most targets
can be observed in one to two observing runs 
(due to fiber collision, there could be a small fraction which is 
left unobserved). For MAIN2, the angular density is about $1050 \deg^{-2}$. 
These can be observed in 5-6 observing runs.

The total number of target is about 2.6 million for the MAIN1 survey, 
and 8.4 million
for the MAIN2 survey. The total fiber time required for completing the survey is
about 0.88 million fiber hours for MAIN1, and 14 million fiber hours 
for MAIN2.  If we assume that all fibers are allocated for
this survey, and for each night the observing time is 6
hours, then MAIN1 observation can be completed in 37 clear dark (moonless) 
nights, and MAIN2 in 584 clear dark nights. 
In reality, the automatic survey strategy system will probably assign targets 
of different surveys in the same sky area at the same time, and only 
1/3-1/2 of the nights have good weather. Thus, 
we expect that the MAIN1 sample can be surveyed in the first year, 
while the MAIN2 sample in the next 5-10 years, depending on the weather 
condition and the fraction of fiber time allocated for this survey 
\footnote{This estimate may still be optimistic. One problem of the Xinglong
site is that during the spring to summer time
when much of the SDSS surveyed region can be observed at meridian transit, 
the weather condition is worse than the autumn (H. Wu, 
private communication).}. 

\begin{table}
\caption{\label{tab:hod} HOD parameters for galaxies (reproduced from \citet{ZZW05}).}
\begin{tabular}{c c c c c}
\hline
$M^{\rm max}_{r}$ & $\log_{10}m_{min}$  &  $\log_{10}m_{1}$ & $\alpha$ & \\
\hline
-22.0  & 13.91 & 14.92 & 1.43 & \\
-21.5  & 13.27 & 14.60 & 1.94 & \\
-21.0  & 12.72 & 14.09 & 1.39 & \\
-20.5  & 12.30 & 13.67 & 1.21 & \\
-20.0  & 12.01 & 13.42 & 1.16 & \\
-19.5  & 11.76 & 13.15 & 1.13 & \\
-19.0  & 11.59 & 12.94 & 1.08 & \\
-18.5  & 11.44 & 12.77 & 1.01 & \\
-18.0  & 11.27 & 12.57 & 0.92 & \\
\hline
\end{tabular}\\
\end{table}

\begin{figure}
\includegraphics[height=0.45\textwidth]{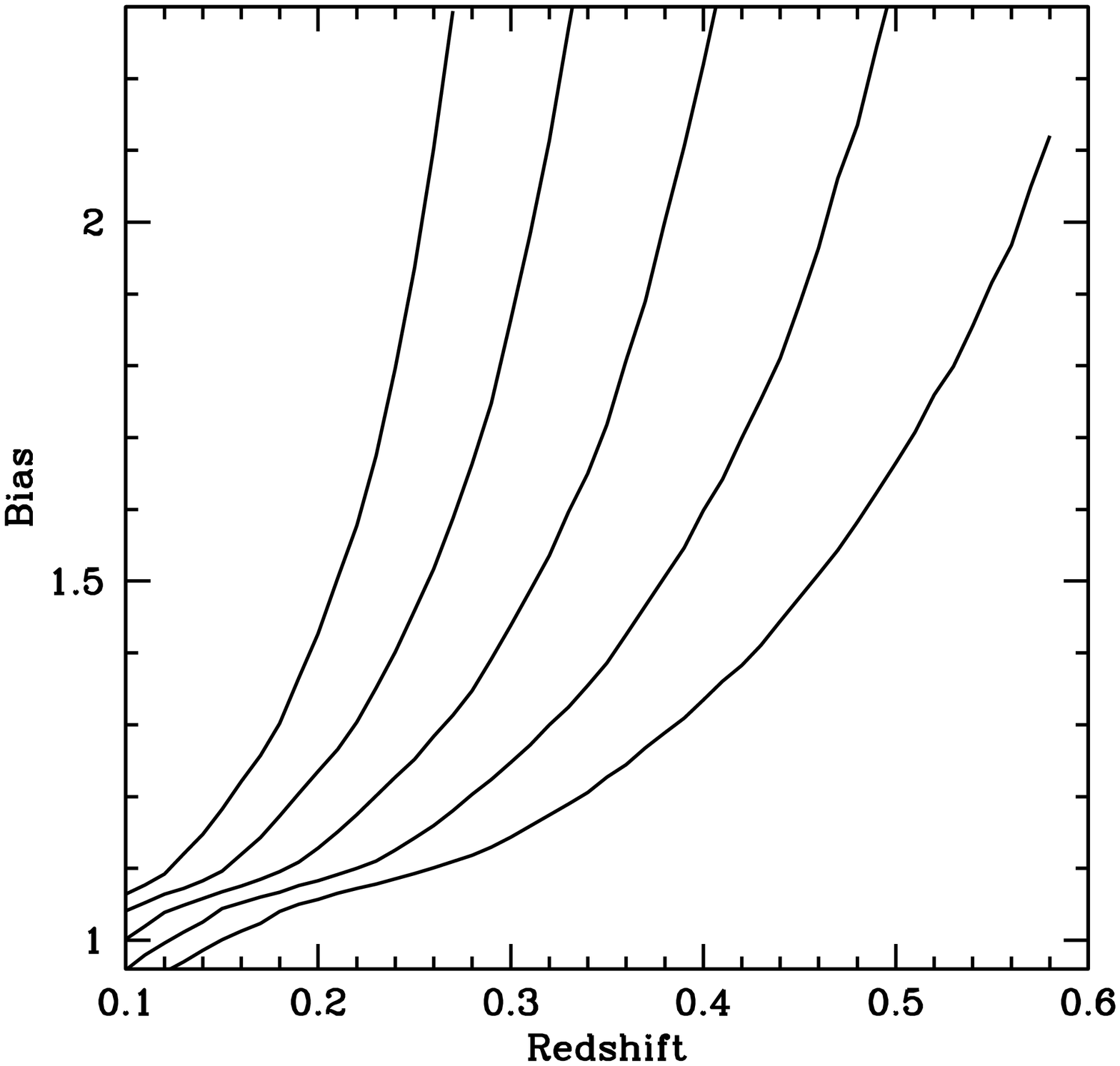}
\caption{\label{fig:bias_main}The clustering bias of the main sample as a 
function of redshift. Curves
from the top down correspond to the magnitude limit
of 17.8, 18.3, 18.8, 19.2, 19.8.}
\end{figure}

We can estimate the bias of the main sample using the halo model as outlined
in \S 2.3. The HOD parameter for all galaxy types have
been measured for the SDSS main sample DR2 \citep{ZZW05}, which consists
of 204,584 galaxies over $2,947 \deg^2$ with apparent luminosity
$10.5<r<17.77$ and redshifts extended to $z\sim 0.2$. The 
mean number of galaxies above 
limiting luminosity $L_{lim}$ in halos of mass $m$ 
is parameterized as
\begin{equation}
\label{eq:HOD}
\langle N(m,L>L_{lim})\rangle =\Theta(m-m_{\rm min})[1+(m/m_1)^\alpha],
\end{equation}
where $\Theta(x)$ is the Heaviside step function, and the HOD 
parameters $\theta \equiv \{m_{min}, m_1, \alpha \}$ are all functions 
of the limiting luminosity $L_{lim}$ and redshift $z$, 
their measured values are given 
in Table 3 of \cite{ZZW05}. However, \cite{ZZW05} gives only the measured
value at $z_{m}\sim 0.1$, at higher redshift the data is not 
available. Since the LAMOST galaxy sample would have 
much higher average redshift, evolution of HOD parameters should be considered.
We have adopted the simple {\it ansatz} that the HOD parameters are functions 
of one parameter $\lambda$ only, $\lambda$ is a function of both the 
actual limiting luminosity $L$ and the redshift $z$. 
At $z=z_m$, $\lambda(L,z_m)=L$, 
and $\theta[\lambda(L,z)]=\theta(L,z_m)$ as given in 
\cite{ZZW05}. With this {\it ansatz}, 
in order to know $\theta(L,z)$ at a higher $z$, 
we only need to know the corresponding
$\lambda(L,z)$. To obtain $\lambda$, we calculate the galaxy 
density at $z$ using the HOD parameters obtained 
with a trial value of $\lambda$, 
compare it with the density integrated from the measured luminosity function 
$\phi(L,z)$ given in  \cite{B03}, 
\begin{equation}
  \int_0^{\infty} dm \frac{dn}{dm} \langle N(m,L>L_{lim}, \lambda)\rangle
= \int_{L_{lim}}^{\infty} \phi(L,z) dL,
\end{equation}
where $\langle N(m,L>L_{lim}, \lambda)\rangle$ is given in eq.~\ref{eq:HOD}.
The value of $\lambda$ is adjusted so that the galaxy densities calculated 
with the HOD  matches the observation. Of course, this 
{\it ansatz} may not be true. However, since we match the luminosity 
function at different redshifts, the estimate of bias would not be 
too far off.

By using this method, we
obtain an estimate of the bias at different redshifts. For example, 
for the MAIN1 sample, at $z=0.1$, the bias is $b=1.0$, 
and at $z=0.3$, $b= 1.44$. 
For the MAIN2 sample, $ b = 0.92$ at $z=0.1$, $b=1.14 $ at $z=0.3$, and 
$ b = 1.66$ at $z=0.5$.
The bias as a function redshift is plotted in Fig.~\ref{fig:bias_main}. 
For reference, besides the 
$r=18.8$ and $19.8$ main survey samples, 
we also plotted several samples of different 
magnitude limits.

\begin{figure}
\includegraphics[height=0.4\textwidth]{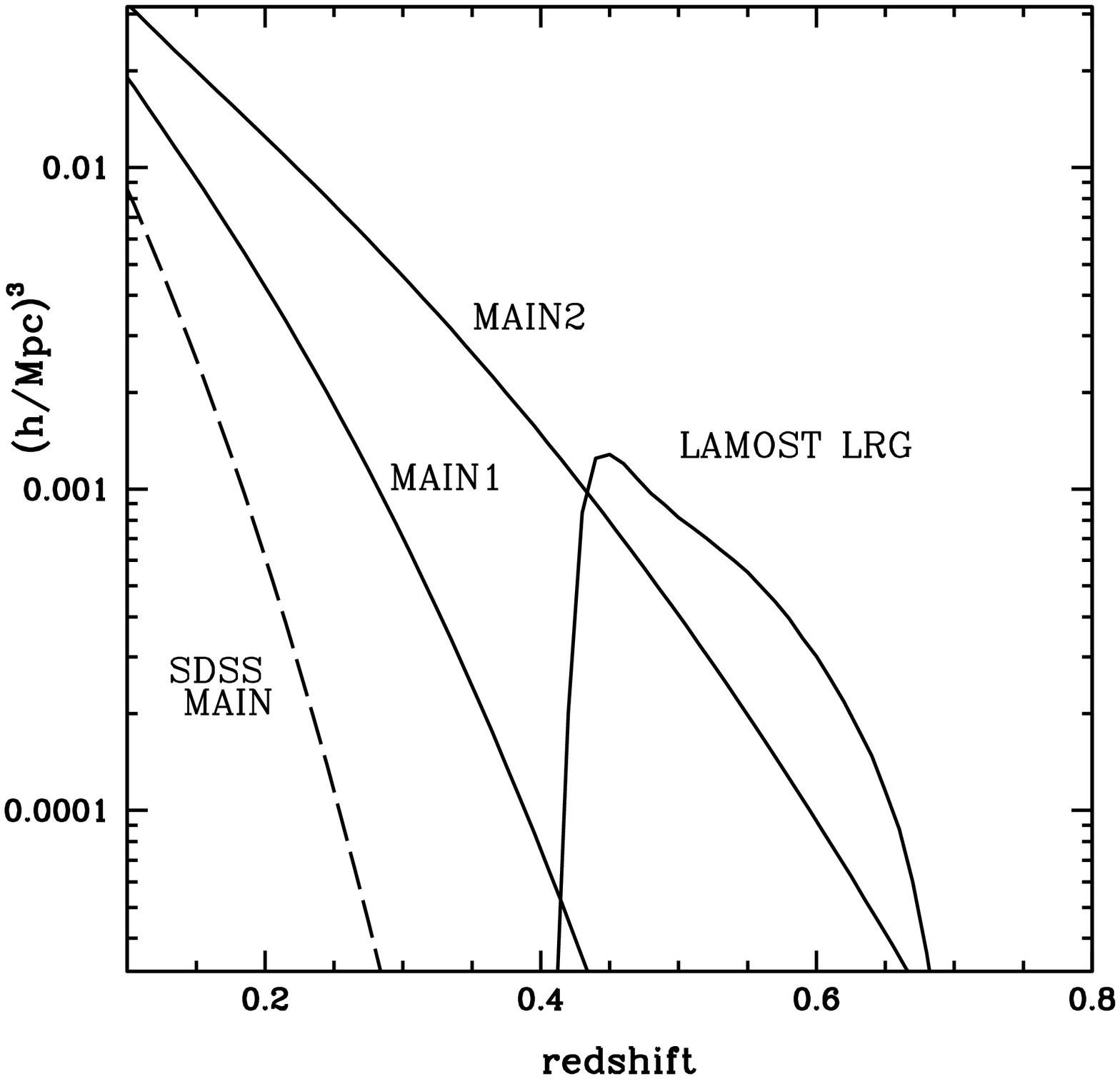}
\caption{\label{fig:dist_galaxy_den} The 
comoving number density of galaxy samples.}
\end{figure}

The distributions of galaxies as a function of redshift are 
plotted in Fig.~\ref{fig:dist_galaxy_z}.
The MAIN1 sample peaks at $z \sim 0.2$, but extends beyond $z=0.4$, where
comoving number density decrease to $n(z=0.4)\approx 7\times10^{-5}
 (h/\Mpc)^3$.  The MAIN2 sample has a more extended distribution,
up to $z=0.6$ with the same cut-off density. The corresponding comoving 
density distributions are plotted in Fig.~\ref{fig:dist_galaxy_den}.

\subsection{The LRG survey}

The LRGs are intrinsically brighter and can be selected by their colors.
At present, a catalogue of LRGs have already been compiled from the 
SDSS photometric data,
namely the MegaZ LRG catalogue \citep{C06}. 
This is a catalogue of LRG galaxies selected 
from the SDSS DR4 imaging data, which covers $5914\deg^2$,
with  $17.5<i_{dev}<20.0$. The total number of galaxies in this
catalogue is 1.2 million, and photometric redshifts for these galaxies
have been obtained, with an error of $\sigma_z =0.048$. It is expected
that the sample will be extended to $8000\deg^2$, and a total number of
1.5 million of LRGs. 
The sample selection of LRGs might be further tailored for the needs of LAMOST,
nevertheless it should have a lot in 
common with the MegaZ sample, particularly at $z>0.4$. We therefore
tentatively take the properties of this sample as the LAMOST LRG survey 
sample in our estimation.

We expect that an exposure time of at least 70 minutes is required  
for observing most of these targets. The angular density of the sample is about 
205 per square degree, which is very close to the angular number density of 
the fibers of LAMOST (200). The total observing fiber time required
for completing 8000 square degree area is about 1.9 million fiber 
hours (80 nights).

Note that there is not a unique selection criterion of LRGs, hence the LRGs 
in the MegaZ sample may differ slightly from those already spectroscopically 
observed in the SDSS-II, or those observed in the 2SLAQ survey \citep{CN06}. 
The color cut used by MegaZ is similar to but somewhat 
broader than that used by 2SLAQ. It includes bluer objects, 
which is in turn bluer
than the Cut II of SDSS LRG to accommodate the higher density of 2dF fibers.
\cite{W06} has compared the  selection criterion difference between
SDSS and 2SLAQ for the evolution of luminosity function,
they started from SDSS main sample which consists of all galaxies with
$r<17.77$,  then  evolved these to $z=0.2$ (for SDSS) and $z=0.55$ (for 2SLAQ) 
by interpolating between passive and continous star formation model 
according to the observed $g-i$ color. 
They found that 90\% of the SDSS LRG would be selected by 2SLAQ, while 
only 30\% of the 2SLAQ LRG would be selected by SDSS.
The number distribution of the MegaZ sample (marked LAMOST LRG)
is shown in Fig.~\ref{fig:dist_galaxy_z},
most targets in this catalogue have redshifts between 0.4 and 0.7.
It peaks aroud $z=0.45$. Nearly 20\% of the objects in the 
catalogue \citep{C06} are located 
within the region with
$0.50 < d_\perp < 0.55,~ d_\perp =(r - i) - (g - r) /8.0 $.  
If these objects are removed, then the photometric redshift distributions
of the MegaZ-LRG catalogue and the 2SLAQ evaluation set would become similar 
(c.f. Figure.~12 in \cite{C06}).

To estimate the bias of the Mega-Z sample, we use the HOD parameters
for red galaxies as given in \cite{BZ08}, which is obtained by 
using the observed luminosity function and the luminosity-dependent
clustering of 40696 red galaxies with $0.2<z<1.0$.
The HOD parameters  can be 
approximated as a function of redshift and B-band absolute magnitude 
$M_B + 1.2z$, which is an effective proxy for stellar mass:
\begin{eqnarray}
M_{min}(h_{-1} M_{\odot} )=  10^{11.95} \times 10^{0.4\times 
                 [-19 - (M_B- 5\log h + 1.2z) ] }  \nonumber \\
        +10^{13.70} \times 10^{1.15 \times 
                 [-21 -(M_B-5\log h+1.2z) ] }   
        + 10^{11.85}   \\
M^{'}_1 (h_-1 M_{\odot} )  = 10^{12.70}  
    \times 10^{0.11\times[-17 -(M_B-5\log h+1.2z) ] } \nonumber\\
  + 10^{14.60} \times 10^{0.85\times[-21 -(M_B-5\log h+1.2z) ]  }
\end{eqnarray}
with $M_0 = M_{min}$, $\sigma_{\log M} = 0.3 $ and $\alpha = 1$. 
Since the bands are different and the transformation  between them are
complicated, again, we calculate the bias by tuning the B band 
magnitude threshold to match the space density, then use the formalism
presented in \S 2.3.
We find the biases of the sample are 1.77 at z=0.475, and 2.29 at z=0.625.

\begin{table}
\caption{\label{tab:bias}Bias}
\begin{tabular}{|c  c | c  c|c  c |}
\hline
\multicolumn{2}{|c|}{MegaZ} & \multicolumn{2}{|c|}{MAIN1 } 
   & \multicolumn{2}{|c|}{MAIN2 } \\
  z & $b_g$ &  z & $b_g$ &  z & $b_g$ \\
 \cline{1-6}
    0.475 & 1.77   & 0.10 & 1.00&    0.1 & 0.92\\
    0.625 & 2.29   & 0.3  & 1.44 &   0.3 & 1.14\\
          &       &       &      &   0.5 &  1.66\\
\hline
\multicolumn{2}{|c|}{QSO1} & \multicolumn{2}{|c|}{QSO2} 
   & \multicolumn{2}{|c|}{QSO3} \\
z & $b_g$ &  z & $b_g$ & z & $b_g$\\
\cline{1-6}
   0.575  &  1.26 &   0.530   & 1.23   &   0.53 &    1.24   \\
   0.885  &  1.57 &   0.790   & 1.41   &   0.80  &   1.42    \\
   1.155  &  1.92 &   1.068 &   1.72  &    1.07  &   1.62  \\ 
   1.425  &  2.32 &   1.363 &   2.11  &    1.35  &   1.96  \\
   1.695  &  2.77 &   1.658 &   2.55  &    1.65  &   2.36  \\
   1.965  &  3.26 &   1.953 &   3.04  &    1.95  &   2.79  \\
\hline
\end{tabular}
\end{table}

\subsection{QSO survey}

Finally, we consider a QSO survey. The Ly$\alpha$ forest of the QSO
can be a powerful probe of the baryon acoustic oscillation \citep{ME07}, but
here the QSO itself is used as tracers.
In past surveys such as the
2dF and SDSS, the low number density of QSOs prevents it from 
being a good tracer target for measuring BAO features.
However, with LAMOST the comoving number density can be greater, making
it more useful for such purpose.
Using the 2SLAQ sample \citep{R05}, which 
consists of 5645 quasars down to $g=21.85$ in 
105.7 deg$^2$,  we obtain a luminosity function of the quasars. 
Unlike the Schechter form of galaxies luminosity function, 
the QSO luminosity function is modelled as a double
power law \citep{C04,R05}
\begin{equation}
\Phi (M_g,z) = \frac{\Phi^* } { 10^{0.4(\alpha+1)(M_g-M_g^* ) }
       + 10 ^{0.4(\beta + 1)(M_g- M_g^* )} }
\end{equation}
For estimation we assume here that the evolution with redshift is 
pure luminosity evolution, individual quasars is fainter today than at
z = 2, with the dependence of the characteristic luminosity
described by \citep{R05}  
\begin{equation}
M_g^* (z)  = M_g^* -  2.5(k_1 z + k_2 z^2)
\end{equation}
where $\alpha= -3.28,~ \beta=-1.78,~ M^*=-22.68,~ 
k_1 = 1.37,~ k_2= -0.32,~ \Phi ^*= 5.96\times 10^{-7} \Mpc^{-3}$.
In reality, the evolution of quasar luminosity function may be due to a
combination of decreasing space density and decreasing luminosity of 
individual quasars, but this would not significantly affect our result.
We calculated the QSO K-correction by assuming that the continuum 
$F_{\nu} \propto \nu^{-0.5}$.

\begin{figure}
\includegraphics[height=0.4\textwidth]{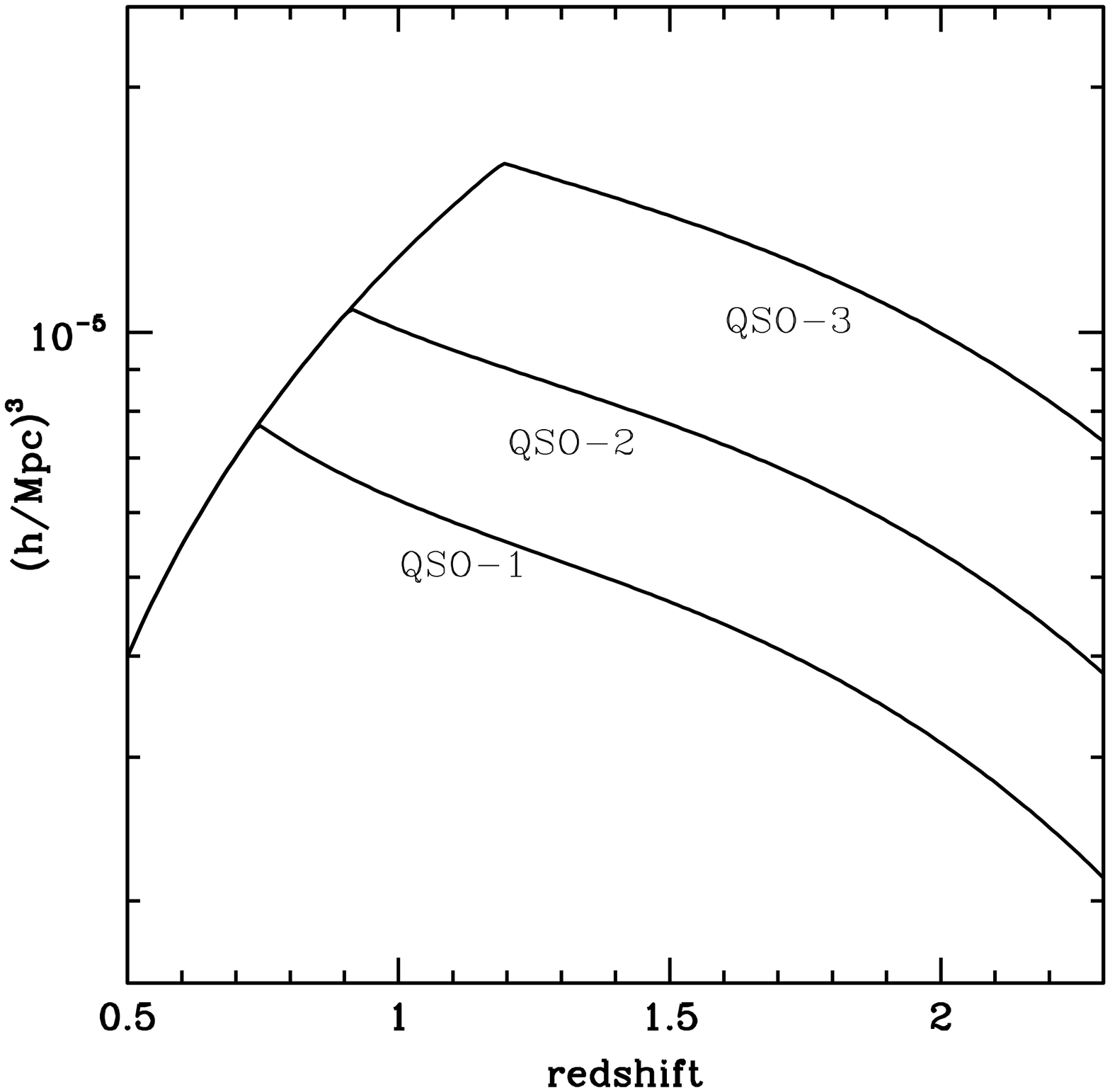}
\caption{\label{fig:den_qso}Comoving number density of the 
QSO samples with QSO1 ($g < 20.5$), QSO2 ($g < 21.0$) and QSO3 
($g < 21.65$).}
\end{figure}

The quasar number density is plotted in Fig.~\ref{fig:den_qso}.
Consider samples with the magnitude limits, 
$g < 20.5$ (QSO1), $g < 21.0$ (QSO2), and  
$g < 21.65$ (QSO3), all between $z=0.4$ and $z=2.1$.
The density of these magnitude limited samples drops at low redshift, 
because there is a lower limit $M_g < -22.5$ on the absolute luminosity for 
quasars (this also reflects a decrease in quasar activity at low redshifts). 
The turning redshift $z_t$ is the place where 
$M_g ( z_t, m_{ g,lim} ) = M_{ g,lim}$, and $ M_{ g, lim} = -22.5$.
They are $z_{t, QSO1}= 0.75, ~ z_{t, QSO2}=0.92 , ~z_{t, QSO3}= 1.20$.

The angular density of the QSO1 is $30 \deg^{-2}$, and the total number is
0.24 million for 8000 $\deg^2$. 
The observing time is about 15 minutes for signal-to-noise ratio
$S/N=4 $, so it requires about 0.06 million fiber hours.
For QSO2 sample, the angular density is $45 \deg^{-2}$, the spectrum can 
be obtained within 30 minutes for the same S/N, and requires 0.18 million
fiber hours. The angular density of QSO3 is $72 \deg^{-2} $, 
the required integration time is 1.5 hours, depending on the sky 
background. 
It needs 0.855 million fiber hours to complete the survey. 

The bias of a QSO sample can be estimated using the method given in \S 2.3.
Alternatively, it can also be measured 
directly from the projected two point correlation function.
\citet{P04} made an measurement with about 14,000 quasars at $0.8<z<2.1$ 
in the 2dF/6dF QSO Redshift Survey.
They found a slightly greater bias (see Table.~\ref{tab:bias}) 
than that of the estimate obtained using the method of \S 2.3. 
The difference is greater at higher redshifts (10\% at $z=1.5$, and 25\% 
at $z=1.9$) \citep{MC06}. Our estimate may be improved by 
using more detailed semi-analytical models, but 
for our purpose this simple estimate should suffice.

\label{lastpage}
\end{document}